\documentclass{article}[11pt]

\usepackage{amsmath,amsfonts,amsthm,tikz,fullpage,hyperref,algorithm,algorithmic,multirow,tabularx}

\newcommand{\INPUT}{\item[{\bf Input:}]}
\newcommand{\OUTPUT}{\item[{\bf Output:}]}
\newtheorem{theorem}{Theorem}[section]
\newtheorem{lemma}[theorem]{Lemma}

\newtheorem{corollary}[theorem]{Corollary}

\newtheorem{definition}[theorem]{Definition}
\newtheorem{example}[theorem]{Example}
\newcommand{\partdiff}[2]{\frac{\partial {#1}}{\partial {#2}}}

\newcommand{\mixdiff}[3]{\frac{\partial^2 {#1}}{{\partial {#2}}{\partial {#3}}}}
\newcommand{\E}[1]{{\bf{E}}\left[#1\right]}
\newcommand{\EE}[2]{{\bf{E}}_{#1}\left[#2\right]}

\renewenvironment{proof}{\noindent{\bf Proof}:~}{$\hfill \Box$\\}

\newcommand{\T}[4] {\begin{tabular}{|c|c|}
\hline
#1 & #2 \\
\hline
#3 & #4 \\
\hline
\end{tabular}}

\newcommand{\TD}[4] {\begin{tabular}{r|c|c|}
\multicolumn{1}{r}{}
 & \multicolumn{1}{c}{$C$}
 & \multicolumn{1}{c}{$\overline{C}$} \\
\cline{2-3}
$A$ & #1 & #2 \\
\cline{2-3}
$B$ & #3 & #4 \\
\cline{2-3}
\end{tabular}}

\newcommand{\TTR}[5]{\begin{tabular}{|c|c|l}
\cline{1-2}
#1 & #2 \\
\cline{1-2}
\multirow{2}{*}{#3} & {\small{#4}} \\
\cline{2-2}
& {\small{#5}} \\
\cline{1-2}
\end{tabular}}

\newcommand{\TTL}[5]{\begin{tabular}{|c|c|}
\cline{1-2}
 #1 & #2 \\
\cline{1-2}
{\small{#3}} & \multirow{2}{*}{#5}  \\
\cline{1-1}
 {\small{#4}} &  \\
\cline{1-2}
\end{tabular}}

\def\b1{{\bf 1}}
\def\be{{\bf e}}
\def\bx{{\bf x}}
\def\by{{\bf y}}

\def\bv{{\bf v}}
\def\RR{{\mathbb R}}
\def\ZZ{{\mathbb Z}}
\def\cM{{\cal M}}
\def\cI{{\cal I}}
\def\cB{{\cal B}}
\def\cF{{\cal F}}
\def\cG{{\mathcal{G}}}
\def\eps {\epsilon}
\def\max{{\rm{max}}}
\def\sup{{\rm{sup}}}
\def\G{{G}}

\begin{document}

\title{Submodular Maximization by Simulated Annealing}

\author{
Shayan Oveis Gharan\thanks{Stanford University, Stanford, CA; {\tt shayan@stanford.edu};
this work was done partly while the author was at IBM Almaden Research Center, San Jose, CA.}
\and Jan Vondr\'ak\thanks{IBM Almaden Research Center, San Jose, CA; {\tt jvondrak@us.ibm.com}}
}

\maketitle

\begin{abstract}
We consider the problem of maximizing a nonnegative (possibly non-monotone)
submodular set function with or without constraints. Feige et al. \cite{FMV07}
showed a $2/5$-approximation for the unconstrained problem and also proved that
no approximation better than $1/2$ is possible in the value oracle model.
Constant-factor approximation was also given for submodular maximization subject
to a matroid independence constraint  (a factor of $0.309$ \cite{Vondrak09})
and for submodular maximization subject to a matroid base constraint,
provided that the fractional base packing number is at least $2$
(a $1/4$-approximation \cite{Vondrak09}).

In this paper, we propose a new algorithm for submodular maximization which is based
on the idea of {\em simulated annealing}. We prove that this algorithm achieves
improved approximation for two problems:
a $0.41$-approximation for unconstrained submodular maximization,
and a $0.325$-approximation for submodular maximization subject to a matroid
independence constraint.

On the hardness side, we show that in the value oracle model it is impossible
to achieve a $0.478$-approximation for submodular maximization
subject to a matroid independence constraint, or a $0.394$-approximation
subject to a matroid base constraint in matroids with two disjoint bases.
Even for the special case of cardinality constraint, we prove it is impossible to
achieve a $0.491$-approximation.
(Previously it was conceivable that a $1/2$-approximation exists for these problems.)
It is still an open question whether a $1/2$-approximation is possible
for unconstrained submodular maximization.
\end{abstract}

\thispagestyle{empty}
\newpage
\setcounter{page}{1}

\section{Introduction}

A function $f: 2^X \rightarrow \RR$ is called {\em submodular} if for any $S, T \subseteq X$,
$ f(S \cup T) + f(S \cap T) \leq f(S) + f(T).$ In this paper,
we consider the problem of {\em maximizing a nonnegative submodular function}.
This means, given a submodular function $f: 2^X \rightarrow \RR_+$, find
a set $S \subseteq X$ (possibly under some constraints) maximizing $f(S)$.
We assume a {\em value oracle} access to the submodular function; i.e., for a given set $S$,
the algorithm can query an oracle to find its value $f(S)$.

\paragraph{Background.}
Submodular functions have been studied for a long time in the context of combinatorial
optimization.
Lov\'asz in his seminal paper \cite{Lovasz83} discussed various properties
of submodular functions and noted that they exhibit certain properties
reminiscent of convex functions - namely the fact that a naturally defined extension
of a submodular function to a continuous function (the "Lov\'asz extension") is convex. 
This point of view explains why submodular functions can be {\em minimized}
efficiently \cite{GLS81,FFI00,Schrijver00}.

On the other hand, submodular functions also exhibit properties closer to concavity,
for example a function $f(S) = \phi(|S|)$ is submodular if and only if $\phi$ is concave.
However, the problem of {\em maximizing} a submodular function captures
problems such as Max Cut \cite{GW95} and Max $k$-cover \cite{Feige98} which are NP-hard.
Hence, we cannot expect to maximize a submodular function exactly;
still, the structure of a submodular functions (in particular, the ``concave aspect''
of submodularity) makes it possible to achieve non-trivial results for maximization problems.
Instead of the Lov\'asz extension, the construct which turns out to be useful for maximization
problems is the {\em multilinear extension}, introduced in \cite{CCPV07}. 
This extension has been used to design an optimal $(1-1/e)$-approximation for the
problem of maximizing a monotone submodular function subject to a matroid independence
constraint \cite{Vondrak08,CCPV09}, improving the greedy $1/2$-approximation of
Fisher, Nemhauser and Wolsey \cite{NWF78II}.
In contrast to the Lov\'asz extension,
the multilinear extension captures the concave as well as convex aspects of submodularity.
A number of improved results followed for maximizing monotone submodular functions subject
to various constraints \cite{KST09,LMNS09,LSV09,CVZ10}.

This paper is concerned with submodular functions which are not necessarily monotone.
We only assume that the function is nonnegative.\footnote{For submodular
functions without any restrictions, verifying whether the maximum
of the function is greater than zero or not requires exponentially many queries.
Thus, no approximation algorithm can be found for this problem.}
\iffalse 
In this setting, the maximization problem is computationally hard even without any constraint,
since it captures for example the Max Cut problem.
Typical examples of such a problem are Max Cut and Max Directed Cut.
Here, the best approximation factors have been achieved using
semidefinite programming: $0.878$ for Max Cut~\cite{GW95} and
$0.874$ for Max Di-Cut~\cite{FG95,LLZ02}. The approximation factor for
Max Cut has been proved optimal, assuming the Unique Games
Conjecture~\cite{KKMO04,MOO05}. An algorithm for Max Cut based on eigenvalue computations
rather than SDP is also known to beat the factor of $1/2$ \cite{Trevisan08}.
However, purely combinatorial algorithms for Max Cut and Max Di-Cut are known to
achieve only a $\frac12$-approximation \cite{HZ01}.
\fi
The problem of maximizing a nonnegative submodular function has been studied
in the operations research community, with many heuristic solutions proposed:
data-correcting search methods~\cite{GSTT99,GTT99,K89}, accelatered greedy algorithms~\cite{RS89},
and polyhedral algorithms~\cite{LNW96}.
The first algorithms with provable performace guarantees for this problem were given 
by Feige, Mirrokni and Vondr\'ak \cite{FMV07}. They presented several algorithms achieving
constant-factor approximation, the best approximation factor being $2/5$
 (by a randomized local search algorithm).
They also proved that a better than $1/2$ approximation for submodular maximization
would require exponentially many queries in the value oracle model.
This is true even for symmetric submodular functions, in which case a
$1/2$-approximation is easy to achieve \cite{FMV07}.

Recently, approximation algorithms have been designed for nonnegative
submodular maximization subject to various constraints \cite{LMNS09,LSV09,Vondrak09,GRST10}.
(Submodular minimization subject to additional constraints has been
also studied \cite{SF08,GKTW09,IN09}.)
The results most relevant to this work are that a nonnegative submodular functions can be
maximized subject to a matroid independence constraint within a factor of $0.309$,
while a better than $1/2$-approximation is impossible \cite{Vondrak09},
and there is $\frac12 (1-\frac{1}{\nu}-o(1))$-approximation subject to
a matroid base constraint for matroids of fractional
base packing number at least $\nu \in [1,2]$, while a better
than $(1-\frac{1}{\nu})$-approximation in this setting is impossible \cite{Vondrak09}.
For explicitly represented instances of unconstrained submodular maximization,
Austrin \cite{Austrin10} recently proved that assuming the Unique Games Conjecture,
the problem is NP-hard to approximate within a factor of $0.695$. 
%Even in this setting, the best known approximation was a $2/5$ factor \cite{FMV07}.

\paragraph{Our results.}
In this paper, we propose a new algorithm for submodular maximization, using
the concept of {\em simulated annealing}. The main idea is to perform a local search
under a certain amount of random noise which gradually decreases to zero.
This helps avoid bad local optima at the beginning,
and provides gradually more and more refined local search towards the end.
Algorithms of this type have been widely employed for difficult optimization problems,
but notoriously difficult to analyze.

We prove that the simulated annealing algorithm achieves a $0.41$-approximation for the maximization
of any nonnegative submodular function without constraints, improving upon the previously known
$0.4$-approximation \cite{FMV07}. (Although our initial hope was that this algorithm
might achieve a $1/2$-approximation, we found an example where it
achieves only a factor of $17/35 \simeq 0.486$; see Appendix~\ref{app:tightexample}.)
We also prove that a similar algorithm achieves a $0.325$-approximation for the
maximization of a nonnegative submodular function subject to a matroid independence
constraint (improving the previously known factor of $0.309$ \cite{Vondrak09}).

On the hardness side, we show the following results in the value oracle model:
For submodular maximization under a matroid base constraint, 
it is impossible to achieve a $0.394$-approximation even in the special case
when the matroid contains two disjoint bases.
For maximizing a nonnegative submodular function subject to a matroid independence constraint,
we prove it is impossible to achieve a $0.478$-approximation.
For the special case of a cardinality constraint ($\max \{f(S): |S| \leq k\}$
or $\max \{f(S): |S|=k\}$), we prove a hardness threshold of $0.491$.
We remark that only a hardness
of $(1/2+\epsilon)$-approximation was known for all these problems prior to this work.
For matroids of fractional base packing number $\nu = k/(k-1), k \in \ZZ$,
we show that submodular maximization subject to a matroid base constraint
does not admit a $(1-e^{-1/k}+\epsilon)$-approximation for any $\epsilon>0$,
improving the previously known threshold of $1/k+\epsilon$ \cite{Vondrak09}. 
These results rely on the notion of a {\em symmetry gap} and the hardness construction of
\cite{Vondrak09}.

%The main remaining question is whether a $1/2$-approximation is possible
%for unconstrained submodular maximization.

\begin{figure}[here]
$\begin{array}{|| c || c | c | c | c ||} \hline
\mbox{Problem} & \mbox{Prior approximation} & \mbox{New approximation}
 & \mbox{New hardness} & \mbox{Prior hardness}  \\
\hline \hline
\max \{f(S): S \subseteq X \} & 0.4 & 0.41 & - & 0.5 \\
\hline
\max \{f(S): |S| \leq k \} & 0.309 & 0.325 & 0.491 & 0.5 \\
\hline
\max \{f(S): |S|=k \} & 0.25 & - & 0.491 & 0.5 \\
\hline
\max \{f(S): S \in \cI \} & 0.309 & 0.325 & 0.478 & 0.5 \\
\hline
\max \{f(S): S \in \cB \}^* & 0.25 & - & 0.394 & 0.5 \\
\hline
\end{array} $
\caption{Summary of results: $f(S)$ is nonnegative submodular,
$\cI$ denotes independent sets in a matroid, and
$\cB$ bases in a matroid. $^*$ - in this line we assume the case
where the matroid contains two disjoint bases.
The hardness results hold in the value oracle model.}
\end{figure}

The rest of the paper is organized as follows. In Section~\ref{sec:prelims},
we discuss the notions of multilinear relaxation and simulated annealing,
which form the basis of our algorithms. In Section~\ref{sec:unconstrained},
we describe and analyze our $0.41$-approximation for unconstrained submodular
maximization. In Section~\ref{sec:matroid-constraint}, we describe
our $0.325$-approximation for submodular maximization subject to a matroid
independence constraint. In Section~\ref{sec:hardness},
we present our hardness results.
Many details are deferred to the appendix.

\section{Preliminaries}
\label{sec:prelims}

Our algorithm combines the following two concepts. The first one is
{\em multilinear relaxation}, which has recently proved to be
very useful for optimization problems involving submodular functions
(see \cite{CCPV07,Vondrak08,CCPV09,KST09,LMNS09,Vondrak09}).
The second concept is {\em simulated annealing},
which has been used successfully by practitioners dealing with difficult optimization problems.
Simulated annealing provides good results in many practical scenarios, but typically
eludes rigorous analysis (with several exceptions in the literature:
see e.g. \cite{BT93} for general convergence results, 
\cite{LV03,KV06} for applications to volumes estimation and optimization over convex bodies,
and \cite{SVV07,BSVV08} for applications to counting problems).

%In this work, we provide provable guarantees for our "simulated annealing" algorithm
%for submodular maximization, improving strictly upon the previously
%best known algorithms.

\medskip
\noindent{\bf Multilinear relaxation.}
Consider a submodular function $f:2^X \rightarrow \RR_+$. We define a continuous function
$F:[0,1]^X \rightarrow \RR_+$ as follows: For $\bx \in [0,1]^X$, let $R \subseteq X$ be a random
set which contains each element $i$ independently with probability $x_i$. Then we define
$$ F(\bx) := \E{f(R)} = \sum_{S \subseteq X} f(S) \prod_{i \in S} x_i \prod_{j \notin S}
 (1-x_j).$$
This is the unique multilinear polynomial in $x_1,\ldots,x_n$ which coincides with $f(S)$
on the points $\bx \in \{0,1\}^X$ (we identify such points with subsets $S \subseteq X$ in a natural way).
Instead of the discrete optimization problem
$ \max \{f(S): S \in \cF \} $
where $\cF \subseteq 2^X$ is the family of feasible sets,
we consider a continuous optimization problem
$ \max \{F(x): x \in P(\cF) \} $
where $P(\cF) = \mbox{conv}(\{\b1_S: S \in \cF\})$ is the polytope associated with $\cF$.
It is known due to \cite{CCPV07,CCPV09,Vondrak09} that any fractional solution 
$\bx \in P(\cF)$ where $\cF$ are either all subsets, or independent sets in a matroid,
or matroid bases, can be rounded to an integral solution $S \in \cF$ such that
 $f(S) \geq F(\bx)$. 
Our algorithm can be seen as a new way of approximately solving the relaxed
problem $\max \{F(\bx): \bx \in P(\cF) \}$.

\iffalse
\begin{lemma}
\label{lem:pipage-rounding}
For any matroid $\cM = (X,\cI)$, let $P$ be either the matroid polytope (convex hull of independent sets)
or the matroid base polytope (convex hull of bases). Then there is an efficient randomized
rounding technique which for any point $x \in P$ finds a random vertex $\b1_R$ of $P$ such that
$\E{\b1_R} = x$, and for any submodular function $f:2^X \rightarrow \RR$ and its multilinear extension $F$,
$$ \E{f(R)} \geq F(x).$$
\end{lemma}
\fi

\medskip
\noindent{\bf Simulated annealing.}
The idea of simulated annealing comes from physical processes such as gradual cooling of molten metals,
whose goal is to achieve the state of lowest possible energy. The process starts at a high temperature
and gradually cools down to a "frozen state". The main idea behind gradual cooling is that
while it is natural for a physical system to seek a state of minimum energy,
this is true only in a local sense - the system does not have any knowledge of the global structure
of the search space. Thus a low-temperature system would simply find a local optimum and get stuck there,
which might be suboptimal. Starting the process at a high temperature means that there is more randomness
in the behavior of the system. This gives the system more freedom to explore the search space,
escape from bad local optima, and converge faster to a better solution.
We pursue a similar strategy here.

We should remark that our algorithm is somewhat different from a direct interpretation of
simulated annealing. In simulated annealing, the system would typically evolve as a random walk,
with sensitivity to the objective function depending on the current temperature. Here,
we adopt a simplistic interpretation of temperature as follows.
Given a set $A \subset X$ and $t \in [0,1]$, we define a probability distribution
${\cal R}_t(A)$ by starting from $A$ and adding/removing each element independently with probability $t$.
Instead of the objective function evaluated on $A$, we consider the expectation over
the distribution ${\cal R}_t(A)$.
This corresponds to the {\em noise operator} used in the analysis of boolean functions,
which was implicitly also used in the $2/5$-approximation algorithm of \cite{FMV07}.
Observe that $\E{f({\cal R}_t(A))} = F((1-t) \b1_A + t \b1_{\overline{A}})$,
where $F$ is the multilinear extension of $f$.
The new idea here is that the parameter $t$ plays a role similar to temperature
- e.g., $t=1/2$ means that ${\cal R}_t(A)$ is uniformly random regardless of $A$
("infinite temperature" in physics), while $t=0$ means that there are no fluctuations
present at all ("absolute zero").

We use this interpretation to design an algorithm inspired by simulated annealing:
Starting from $t=1/2$,
we perform local search on $A$ in order to maximize $\E{f({\cal R}_t(A))}$.
Note that for $t=1/2$ this function does not depend on $A$ at all,
and hence any solution is a local optimum.
Then we start gradually decreasing $t$,
while simultaneously running a local search with respect to $\E{f({\cal R}_t(A))}$.
%The rate of decrease in $t$ is so slow that we can assume a local optimum
%is found before we move to a smaller value of $t$.
Eventually, we reach $t=0$ where
the algorithm degenerates to a traditional local search and returns an
(approximate) local optimum.

We emphasize that we maintain the solution generated by previous stages
of the algorithm, as opposed to running a separate local search for each value of $t$. 
This is also used in the analysis, whose main point is to estimate
how the solution improves as a function of $t$.
It is not a coincidence that the approximation provided by our algorithm
is a (slight) improvement over previous algorithms. Our algorithm can be viewed
as a dynamic process which at each fixed temperature $t$ corresponds to a certain
variant of a previous algorithm. We prove that the performance of the simulated annealing
process is described by a differential equation, whose initial condition can be
related to the performance of a previously known algorithm. Hence the fact that an improvement
can be achieved follows from the fact that the differential equation yields
a positive drift at the initial point. The exact quantitative improvement depends
on the solution of the differential equation, which we also present in this work.

\medskip
\noindent{\bf Notation.}
In this paper, we denote vectors consistently in boldface: for example $\bx, \by \in [0,1]^n$.
The coordinates of $\bx$ are denoted by $x_1,\ldots,x_n$. Subscripts next to a boldface
symbol, such as $\bx_0, \bx_1$, denote different vectors. In particular, we use the notation
$\bx_p(A)$ to denote a vector with coordinates $x_i = p$ for $i \in A$ and $x_i = 1-p$
for $i \notin A$. 
In addition, we use the following notation to denote the value of
certain fractional solutions:
\vspace{-5pt}
$$ \TD{$p$}{$p'$}{$q$}{$q'$} := F(p \b1_{A \cap C} + p' \b1_{A \setminus C}
 + q \b1_{B \cap C} + q' \b1_{B \setminus C}).$$
For example, if $p=p'$ and $q=q'=1-p$, the diagram would represent $F(\bx_p(A))$.
Typically, $A$ will be our current solution, and $C$ an optimal solution.
Later we omit the symbols $A,B,C,\overline{C}$ from the diagram.

%%%%%%%%%%%%%%%%%%%%%%%%%%%%%%%%%%%%%%%%%%%%%%%%%
%%%%%%%%%%%%%%%%%%%%%%%%%%%%%%%%%%%%%%%%%%%%%%%%SECTION3
%

\section{Unconstrained Submodular Maximization}
\label{sec:unconstrained}

Let us describe our algorithm for unconstrained submodular maximization.
We use a parameter $p \in [\frac12,1]$, which is related to the ``temperature''
discussed above by $p = 1-t$. We also use a fixed discretization parameter $\delta = 1/n^3$.

\begin{algorithm}
\caption{Simulated Annealing Algorithm For Submodular Maximization}
\label{alg:simulated_unconstrained}
\begin{algorithmic}[1]
\INPUT A submodular function $f:2^X \rightarrow \RR_+$.
\OUTPUT A subset $A \subseteq X$ satisfying $f(A) \geq 0.41 \cdot \max\{f(S): S \subseteq X\}$.
\STATE {\bf{Define}} $\bx_p(A) = p \b1_{A} + (1-p) \b1_{\overline{A}}$.
%\STATE {\bf{Define}} $\Phi_p(A) = F(\bx_p(A)) = \E{f({\cal R}_{1-p}(A))}=\T{p}{p}{1-p}{1-p}$.
\STATE $A \leftarrow \emptyset$.
\FOR {$p \leftarrow 1/2;~ p<1;~ p \leftarrow p+\delta$}
\WHILE {there exists $i \in X$ such that $F(\bx_p(A \Delta \{i\})) > F(\bx_p(A))$}
\STATE  $A \leftarrow A \Delta \{i\}$ 
\ENDWHILE
\ENDFOR
\RETURN the best solution among all sets $A$ and $\overline{A}$ encountered by the algorithm.
\end{algorithmic}
\end{algorithm}

We remark that this algorithm would not run in polynomial time, due to the complexity of finding
a local optimum in Step 4-6. This can be fixed by standard techniques (as in
 \cite{FMV07,LMNS09,LSV09,Vondrak09}), by stopping when the conditions
of local optimality are satisfied with sufficient accuracy. 
We also assume that we can evaluate the multilinear extension $F$, which can be
done within a certain desired accuracy by random sampling.
Since the analysis of the algorithm is already quite technical, we ignore these issues
in this extended abstract and assume instead that a true local optimum is found in Step 4-6.

\begin{theorem}
\label{thm:0.41-approx}
For any submodular function $f:2^X \rightarrow \RR_+$, Algorithm \ref{alg:simulated_unconstrained} returns with
high probability a solution of value
at least $0.41 \cdot OPT$ where $OPT = \max_{S \subseteq X} f(S)$.
\end{theorem}
In Theorem \ref{thm:tightexample} we also show that Algorithm \ref{alg:simulated_unconstrained}
does not achieve any factor better than $17/35 \simeq 0.486$.
First, let us give an overview of our approach
and compare it to the analysis of the $2/5$-approximation in \cite{FMV07}.
The algorithm of \cite{FMV07}
can be viewed in our framework as follows: for a fixed value of $p$, it performs local search
over points of the form $\bx_p(A)$, with respect to element swaps in $A$,
and returns a locally optimal solution. Using the conditions of local optimality,
$F(\bx_p(A))$ can be compared to the global optimum.
Here, we observe the following additional property of a local optimum.
If $\bx_p(A)$ is a local optimum
with respect to element swaps in $A$, then slightly increasing $p$ cannot decrease the value
of $F(\bx_p(A))$. 
%More formally, $\partdiff{}{p} F(\bx_p(A)) \geq 0$. 
During the local search stage, the value cannot decrease either,
so in fact the value of $F(\bx_p(A))$ is non-decreasing throughout the algorithm.
Moreover, we can derive bounds on $\partdiff{}{p} F(\bx_p(A))$ depending
on the value of the current solution. Consequently,
unless the current solution is already valuable enough, we can conclude that an improvement
can be achieved by increasing $p$.
This leads to a differential equation whose solution implies Theorem~\ref{thm:0.41-approx}.

We proceed slowly and first prove the basic fact that if $\bx_p(A)$ is
a local optimum for a fixed $p$, we cannot lose by increasing $p$ slightly.
This is intuitive, because the gradient $\nabla F$ at $\bx_p(A)$
must be pointing away from the center of the cube $[0,1]^X$, or else we could gain
by a local step.

\begin{lemma}
\label{lem:positive-drift}
Let $p \in [\frac12,1]$ and suppose $\bx_p(A)$ is a local optimum in the sense that
$F(\bx_p(A \Delta \{i\})) \leq F(\bx_p(A))$ for all $i$. Then
\begin{itemize}
\item $\partdiff{F}{x_i} \geq 0$ if $i \in A$,
and $\partdiff{F}{x_i} \leq 0$ if $i \notin A$,
\item $\partdiff{}{p} F(\bx_p(A))
 = \sum_{i \in A} \partdiff{F}{x_i} - \sum_{i \notin A} \partdiff{F}{x_i} \geq 0$.
\end{itemize}
\end{lemma}

\begin{proof}
We assume that flipping the membership of element $i$ in $A$ can only decrease the value
of $F(\bx_p(A))$. The effect of this local step on $\bx_p(A)$ is that the value of the $i$-th coordinate
changes from $p$ to $1-p$ or vice versa (depending on whether $i$ is in $A$ or not).
Since $F$ is linear when only one coordinate is being changed, this implies
$\partdiff{F}{x_i} \geq 0$ if $i \in A$, and $\partdiff{F}{x_i} \leq 0$ if $i \notin A$.
By the chain rule, we have
$$ \partdiff{F(\bx_p(A))}{p} = \sum_{i=1}^{n} \partdiff{F}{x_i} \frac{d (\bx_p(A))_i}{dp}.$$
Since $(\bx_p(A))_i = p$ if $i \in A$ and $1-p$ otherwise, we get
$ \partdiff{F(\bx_p(A))}{p} = \sum_{i \in A} \partdiff{F}{x_i} -
 \sum_{i \notin A} \partdiff{F}{x_i} \geq 0 $
using the conditions above.
\end{proof}

%This implies that the value of the current solution can only improve at any point.
%However, we need a stronger quantitative estimate on $\partdiff{}{p} F(\bx_p(A))$
%in order to prove that a non-trivial improvement can be achieved.
In the next lemma, we prove a stronger bound on the derivative $\partdiff{}{p} F(\bx_p(A))$
which will be our main tool in proving Theorem \ref{thm:0.41-approx}.
This can be combined with the analysis of \cite{FMV07} to achieve a certain improvement.
For instance, \cite{FMV07} implies that if $A$ is a local optimum for $p=2/3$,
we have either $f(\overline{A}) \geq \frac25 OPT$, or $F(\bx_p(A)) \geq \frac25 OPT$. 
Suppose we start our analysis from the point $p = 2/3$. (The algorithm does not need
to be modified, since at $p=2/3$ it finds a local optimum in any case, and this is
sufficient for the analysis.)
We have either $f(\overline{A})>\frac25 OPT$ or $F(\bx_p(A))>\frac25 OPT$,
or else by the following lemma, $\partdiff{}{p} F(\bx_p(A))$ is a constant fraction of $OPT$:
$$ \frac{1}{3} \cdot \partdiff{}{p} F(\bx_p(A)) \geq OPT \, \left(1-\frac{4}{5}
 - \frac{1}{3}\times\frac{2}{5}\right) = \frac{1}{15}OPT.$$
Therefore, in some $\delta$-sized interval, the value of $F(\bx_p(A))$
will increase at a slope proportional to $OPT$. Thus the approximation factor of
 Algorithm \ref{alg:simulated_unconstrained} is strictly greater than $2/5$.
We remark that we use a different starting point to achieve the factor of $0.41$,
and we defer the precise analysis to Appendix~\ref{app:unconstrained}.
%^However, the following lemma is crucial for any variant of our analysis.

\begin{lemma}
\label{lem:drift-bound}
Let $OPT=\max_{S \subseteq X} f(S)$, $p \in [\frac12,1]$ and suppose $\bx_p(A)$ is
a local optimum in the sense that 
$F(\bx_p(A \Delta \{i\})) \leq F(\bx_p(A))$ for all $i$. Then
$$ (1-p) \cdot \partdiff{}{p} F(\bx_p(A)) \geq OPT - 2 F(\bx_p(A)) - (2p-1) f(\overline{A}).$$
\end{lemma}

\begin{proof}
Let $C$ denote an optimal solution, i.e. $f(C) = OPT$.
Let $A$ denote a local optimum with respect to $F(\bx_p(A))$, and $B = \overline{A}$ its complement.
In our notation using diagrams,
$$ F(\bx_p(A)) = F(p \b1_A + (1-p) \b1_B) = \T{p}{p}{1-p}{1-p} $$
The top row is the current solution $A$, the bottom row is its complement $B$,
and the left-hand column is the optimum $C$.
We proceed in two steps. Define 
$$ G(\bx) = (\b1_C - \bx) \cdot \nabla F(\bx) = \sum_{i \in C} (1-x_i) \partdiff{F}{x_i}
 - \sum_{i \notin C} x_i \partdiff{F}{x_i} $$
to denote the derivative of $F$ when moving from $\bx$ towards the actual optimum $\b1_C$.
By Lemma~\ref{lem:positive-drift}, we have
\begin{eqnarray*}
 (1-p) \partdiff{F(\bx_p(A))}{p} & = &
 (1-p) \left( \sum_{i \in A} \partdiff{F}{x_i} - \sum_{i \in B} \partdiff{F}{x_i} \right) \\
  & \geq & (1-p) \left( \sum_{i \in A \cap C} \partdiff{F}{x_i} - \sum_{i \in B \setminus C}
 \partdiff{F}{x_i} \right) - p \left( \sum_{i \in A \setminus C} \partdiff{F}{x_i}
- \sum_{i \in B \cap C}  \partdiff{F}{x_i} \right) = G(\bx_p(A))
%& = & \sum_{i \in C} (1-x_i) \partdiff{F}{x_i}
% - \sum_{i \notin C} x_i \partdiff{F}{x_i} = G(\bx)
\end{eqnarray*}
using the definition of $\bx_p(A)$ and the fact that
$\partdiff{F}{x_i} \geq 0$ for $i \in A \setminus C$ and
$\partdiff{F}{x_i} \leq 0$ for $i \in B \cap C$.
%Finally, note that due to the structure of $\bx = \bx_p(A)$, the last expression is equal to $G(\bx)$.

Next, we use Lemma~\ref{lem:submod-change} to estimate $G(\bx_p(A))$ as follows.
To simplify notation, we denote $\bx_p(A)$ simply by $\bx$. 
If we start from $\bx$ and increase the coordinates in $A \cap C$ by $(1-p)$
and those in $B \cap C$ by $p$, Lemma~\ref{lem:submod-change} says
the value of $F$ will change by 
\begin{equation}
\label{eq:unconstrained_one}
\T{1}{p}{1}{1-p} - \T{p}{p}{1-p}{1-p}
 = F(\bx + (1-p) \b1_{A \cap C} + p \b1_{B \cap C}) - F(\bx) \leq
(1-p) \sum_{i \in A \cap C} \partdiff{F}{x_i} \Big|_\bx 
+ p \sum_{i \in B \cap C} \partdiff{F}{x_i} \Big|_\bx.
\end{equation}
Similarly, if we decrease the coordinates in $A \setminus C$ by $p$ and
those in $B \setminus C$ by $1-p$, the value will change by
\begin{equation}
\label{eq:unconstrained_zero}
 \T{p}{0}{1-p}{0} - \T{p}{p}{1-p}{1-p} 
 = F(\bx - p \b1_{A \setminus C} - (1-p) \b1_{B \setminus C}) - F(\bx)
\leq - p \sum_{i\in A\setminus{C}} \partdiff{F}{x_i} \Big|_\bx
 - (1-p) \sum_{i\in B\setminus{C}} \partdiff{F}{x_i} \Big|_\bx.
\end{equation}
Adding inequalities\eqref{eq:unconstrained_one}, \eqref{eq:unconstrained_zero}
and noting the expression for $G(\bx)$ above, we obtain:
\begin{eqnarray}
\label{eq:prior_differentialeq}
\T{1}{p}{1}{1-p} + \T{p}{0}{1-p}{0} - 2~ \T{p}{p}{1-p}{1-p} & \leq &
% (1-p)\left(\sum_{i\in A\cap {C}}\partdiff{F}{x_i} - \sum_{i\in B\setminus{C}}\partdiff{F}{x_i} \right)
% - p \left(  \sum_{i\in A\setminus {C}}\partdiff{F}{x_i} - \sum_{i\in B\cap{C}}\partdiff{F}{x_i} \right)\\
% & = &
 G(\bx).
\end{eqnarray}
It remains to  relate the LHS of equation \eqref{eq:prior_differentialeq} to the value of $OPT$. We use the "threshold lemma" (see Lemma~\ref{lem:threshold}, and the accompanying example with equation \eqref{eq:threshold_easy}):
%which says that $F(\by) \geq \E{f(T_{>\lambda}(\by))}$,
%where $T_{\lambda}(\by) = \{ i: y_i > \lambda \}$ and $\lambda \in [0,1]$ is uniformly random.
%We apply this lemma to $\by = \bx \wedge \b1_C$ and $\by = \bx \vee \b1_C$, to get the following:
$$ \T{p}{0}{1-p}{0} \geq (1-p) \T{1}{0}{1}{0} + (2p-1) \T{1}{0}{0}{0} + (1-p) \T{0}{0}{0}{0}
 \geq (1-p) OPT + (2p-1) \T{1}{0}{0}{0},$$
$$ \T{1}{p}{1}{1-p} \geq (1-p) \T{1}{1}{1}{1} + (2p-1) \T{1}{1}{1}{0} + (1-p) \T{1}{0}{1}{0}
 \geq (2p-1) \T{1}{1}{1}{0} + (1-p) OPT.$$
Combining these inequalities with (\ref{eq:prior_differentialeq}), we get
$$ G(\bx) \geq 2(1-p) OPT  - 2 ~\T{p}{p}{1-p}{1-p} + (2p-1) \left[~\T{1}{1}{1}{0} + \T{1}{0}{0}{0} ~\right].$$
Recall that $F(\bx) = \T{p}{p}{1-p}{1-p}$. Finally, we add $(2p-1) f(\overline{A})=(2p-1)\T{0}{0}{1}{1}$ to this inequality,
so that we can use submodularity to take advantage
of the last two terms: %$f(A \cup C) + f(A \cap C) + f(\overline{A})
% \geq f(A \cup C) + f(\overline{A} \cup C) \geq f(C) = OPT$. Consequently,
\begin{eqnarray*} G(\bx) + (2p-1) f(\overline{A}) & \geq &
 2(1-p) OPT  - 2 ~\T{p}{p}{1-p}{1-p} + (2p-1) \left[~\T{1}{1}{1}{0} + \T{1}{0}{0}{0} + \T{0}{0}{1}{1} ~\right]\\
 &\geq& 2(1-p) OPT - 2F(\bx_p(A))+ (2p-1)OPT 
  =  OPT - 2 F(\bx_p(A)).
\end{eqnarray*}
\end{proof}

We have proved that unless the current solution is already very valuable, there is a certain
improvement that can be achieved by increasing $p$. The next lemma transforms this statement
into an inequality describing the evolution of the simulated-annealing algorithm.

\begin{lemma}
\label{lem:annealing-dynamics}
Let $A(p)$ denote the local optimum found by the simulated annealing algorithm 
at temperature $t=1-p$,
and let $\Phi(p) = F(\bx_p(A(p)))$ denote its value. Assume also
that for all $p$, we have $f(\overline{A(p)}) \leq \beta$. Then
$$ \frac{1-p}{\delta} (\Phi(p+\delta) - \Phi(p)) \geq 
 (1-2\delta n^2) OPT - 2 \Phi(p) - (2p-1) \beta.$$
\end{lemma}

\begin{proof}
Here we combine the positive drift obtained from decreasing the temperature
(described by Lemma~\ref{lem:drift-bound})
and from local search (which is certainly nonnegative).
Consider the local optimum $A$ obtained at temperature $t = 1-p$.
Its value is $\Phi(p) = F(\bx_p(A))$.
By decreasing temperature by $\delta$, we obtain a solution $\bx_{p+\delta}(A)$,
whose value can be estimated in the first order by the derivative at $p$
(see Lemma~\ref{lem:Taylor} for a precise argument):
$$ F(\bx_{p+\delta}(A)) \geq F(\bx_p(A)) + \delta \partdiff{F(\bx_p(A))}{p}  
 - \delta^2 n^2 \, \sup \Big|\mixdiff{F}{x_i}{x_j}\Big|. $$
This is followed by another local-search stage, in which we obtain a new local optimum $A'$.
In this stage, the value of the objective function cannot decrease, so we have
$ \Phi(p+\delta) = F(\bx_{p+\delta}(A')) \geq F(\bx_{p+\delta}(A))$.
% and consequently
%$$ \Phi(p+\delta) \geq \Phi(p) + \delta \partdiff{F(\bx_p(A))}{p}  - \delta^2 n^2 \, 
% \sup \Big|\mixdiff{F}{x_i}{x_j}\Big|.$$
We have $\sup |\mixdiff{F}{x_i}{x_j}| \leq \max_{S,i,j} |f(S+i+j)-f(S+i)-f(S+j)+f(S)| \leq 2 OPT$.
We also estimate $\partdiff{}{p} F(\bx_p(A))$ using Lemma~\ref{lem:positive-drift}, to obtain
$$ \Phi(p+\delta) \geq F(\bx_{p+\delta}(A)) \geq
 F(\bx_p(A)) + \frac{\delta}{1-p} (OPT - 2 F(\bx_p(A)) - (2p-1) f(\overline{A}))
 - 2 \delta^2 n^2 \, OPT. $$
Finally, we use $f(\overline{A})\leq \beta$ and $F(\bx_p(A)) = \Phi(p)$
to derive the statement of the lemma.
\end{proof}

We only sketch the remainder of the analysis. By taking $\delta \rightarrow 0$,
the statement of
Lemma~\ref{lem:annealing-dynamics} leads naturally to the following differential equation:
$$ (1-p) \Phi'(p) \geq OPT - 2 \Phi(p) - (2p-1) \beta.$$
This equation can be solved analytically. Starting from initial condition $\Phi(p_0) = v_0$,
we get for any $p>p_0$:
$$ \Phi(p) \geq \frac12 (1-\beta) + 2 \beta (1-p) -
 \frac{(1-p)^2}{(1-p_0)^2} \left(\frac12 (1-\beta) + 2 \beta (1-p_0) - v_0 \right).$$ 
Choosing the starting point is a non-trivial issue; for example $p_0 = 1/2$ and $v_0 = 1/4$
(the uniformly random approximation of \cite{FMV07}) does not give any improvement over $2/5$.
It turns out that the best choice is $p_0 = \frac{\sqrt{2}}{1+\sqrt{2}}$,
even though the corresponding value $v_0$ is less than $2/5$.  We prove that we can pick
a value $\beta > 0.41$ such that the solution of the differential equation starting at 
$p_0 = \frac{\sqrt{2}}{1+\sqrt{2}}$ reaches a point $p_1$ such that $\Phi(p_1) \geq \beta$.
Details can be found in Appendix~\ref{app:unconstrained}.

%%%%%%%%%%%%%%%%%%%%%%%%%%%%%%%%%%%%%%%%%%%%%%%%%%%%%%
%%%%%%%%%%%%%%%%%%%%%%%%%%%%%%%%%%%%%%%%%%%%%%%%%%%%%SECTION4

\section{Matroid Independence Constraint}
\label{sec:matroid-constraint}

Let $\cM =(X, \cI)$ be a matroid.
We design an algorithm for the case of submodular maximization subject
to a matroid independence constraint, $\max \{f(S): S \in \cI \}$, as follows. The algorithm uses
fractional local search to solve the optimization problem $\max \{ F(x): x \in P_t(\cM) \}$,
where $P_t(\cM) = P(\cM) \cap [0,t]^X$ is a matroid polytope intersected with a box.
This technique, which has been used already in \cite{Vondrak09},
is combined with a simulated annealing procedure, where the parameter
$t$ is gradually being increased from $0$ to $1$. (The analogy with simulated annealing
is less explicit here; in some sense the system exhibits the most randomness in the middle
of the process, when $t = 1/2$.)  Finally, the fractional solution is rounded
using pipage rounding \cite{CCPV07,Vondrak09}; we omit this stage from the description
of the algorithm.

The main difficulty in designing the algorithm is how to handle the temperature-increasing
step. Contrary to the unconstrained problem, we cannot just increment all variables
which were previously saturated at $x_i=t$, because this might violate the matroid constraint.
Instead, we find a subset of variables that can be increased, by reduction to a bipartite matching
problem. We need the following definitions.

\begin{definition}
\label{def:best-match}
Let $0$ be an extra element not occurring in the ground set $X$, and define formally
$\partdiff{F}{x_0} = 0$. 
For $\bx = \frac{1}{N} \sum_{\ell=1}^{n} \b1_{I_\ell}$ and $i \notin I_\ell$,
we define 
$ b_\ell(i) = \mbox{\em argmin}_{j \in I_\ell \cup \{0\}: I_\ell-j+i \in \cI} \partdiff{F}{x_j}.$
\end{definition}

In other words, $b_\ell(i)$ is the least valuable element which can be exchanged for $i$
in the independent set $I_\ell$. Note that such an element must exist due to matroid axioms.
We also consider $b_\ell(i)=0$ as an option in case $I_\ell + i$ itself is independent.
In the following, $0$ can be thought of as a special ``empty'' element,
and the partial derivative $\partdiff{F}{x_0}$ is considered identically equal to zero.
By definition, we get the following statement.

\begin{lemma}
\label{lem:best-match}
For $b_\ell(i)$ defined as above, we have
$ \partdiff{F}{x_i} - \partdiff{F}{x_{b_\ell(i)}} = \max_{j \in I_\ell \cup \{0\}: I_\ell-j+i \in \cI}
 \left( \partdiff{F}{x_i} - \partdiff{F}{x_j} \right).$
\end{lemma}

The following definition is important for the description of our algorithm.

\begin{definition}
\label{def:ex-graph}
For $\bx = \frac{1}{N} \sum_{\ell=1}^{n} \b1_{I_\ell}$, let $A = \{i: x_i = t\}$.
We define a bipartite ``fractional exchange graph'' $G_x$ on $A \cup [N]$ as follows:
We have an edge $(i,\ell) \in E$, whenever $i \notin I_\ell$.
We define its weight as 
$$ w_{i\ell} = \partdiff{F}{x_i} - \partdiff{F}{x_{b_\ell(i)}}
= \max_{j \in I_\ell \cup \{0\}: I_\ell-j+i \in \cI} \left( \partdiff{F}{x_i} - \partdiff{F}{x_j} \right). $$
\end{definition}

We remark that the vertices of the bipartite exchange graph are not elements of $X$ on both sides, but elements on 
one side and independent sets on the other side.
%but between elements 
%and {\em sets} in which the exchange occurs. For each element/set pair, we consider the most
%profitable exchange, removing the least valuable candidate element (possibly none,
%if $I_\ell+i \in \cI$). 
Now we can describe our algorithm.
% see Figure~\ref{alg:simulated_matroidindependence}.

\begin{algorithm}[htb]
\caption{Simulated Annealing Algorithm for a Matroid Independence Constraint}
\label{alg:simulated_matroidindependence}
\begin{algorithmic}[1]
\INPUT A submodular function $f:2^X \rightarrow \RR_+$ and a matroid $\cM=(X,\cI)$.
\OUTPUT An independent set $A \in \cI$ such that $f(A) \geq 0.325 \cdot \max\{f(S): S\in \cI\}$.
\STATE Let $\bx \leftarrow 0$,  $N\leftarrow n^4$ and $\delta\leftarrow 1/N$.
\STATE {\bf{Define}} $P_t(\cM) = P(\cM) \cap [0,t]^X$
\STATE Maintain a representation of $\bx = \frac{1}{N} \sum_{\ell=1}^{N} \b1_{I_\ell}$ where $I_\ell \in \cI$. % and $A = \{i: x_i = t \}$ .
\FOR {$t \leftarrow 0;~ t<1;~ t\leftarrow t+\delta$}
\WHILE {there is $\bv \in \{ \pm \be_i, \be_i-\be_j: i,j \in X\}$
such that $\bx+\delta \bv \in P_t(\cM)$ and $F(\bx+\delta\bv)>F(\bx)$} 
%\IF  {$\bx - \delta \be_i \in P_t(\cM)$ and $F(\bx - \delta \be_i) > F(\bx)$} 
%\STATE  $\bx \leftarrow \bx - \delta \be_i$, 
%\ELSIF {$\bx + \delta \be_i \in P_t(\cM)$ and $F(\bx + \delta \be_i) > F(\bx)$}
%\STATE $\bx \leftarrow \bx + \delta \be_i$
%\ELSIF {$\bx -\delta \be_i + \delta \be_j \in P_t(\cM)$ and $F(\bx - \delta \be_i + \delta \be_j) > F(\bx)$}
%\STATE $\bx := \bx - \delta \be_i + \delta \be_j$.
\STATE  $\bx := \bx + \delta \bv$ \COMMENT{{\bf{Local search}}}
% \ENDIF
\ENDWHILE
\FOR[{{\bf Complementary solution check}}]{each of the $n$ possible sets $T_{\le\lambda}(\bx) = \{i: x_i \le \lambda\}$}
\STATE Find a local optimum $B \subseteq T_{\le\lambda}(\bx), B \in \cI$ trying to maximize $f(B)$.
\STATE Remember the largest $B$ as  a possible candidate for the output of the algorithm%$f(B)$ is larger than the best set encountered so far, rememeber $B$.
\ENDFOR
\STATE Form the fractional exchange graph (see Definition~\ref{def:ex-graph}) 
and find a max-weight matching $M$. 
\STATE Replace $I_\ell$ by $I_\ell - b_\ell(i) + i$ for each edge $(i,\ell) \in M$, and
update the  point $\bx = \frac{1}{N} \sum_{\ell=1}^{N} \b1_{I_\ell}$.
\COMMENT{{\bf Temperature relaxation}: each coordinate increases by at most
$\delta = 1/N$ and hence $\bx \in P_{t+\delta}(\cM)$.}
\ENDFOR
\RETURN  the best encountered solution.
\end{algorithmic}
\end{algorithm}

\begin{theorem}
\label{thm:0.325-approx}
For any submodular function $f:2^X \rightarrow \RR_+$ and matroid $\cM = (X,\cI)$,
Algorithm \ref{alg:simulated_matroidindependence} returns with high probability
a solution of value at least $0.325 \cdot OPT$ where $OPT = \max_{S \in \cI} f(S)$.
\end{theorem}

Let us point out some differences between the analysis of this algorithm
and the one for unconstrained maximization (Algorithm~\ref{alg:simulated_unconstrained}).
The basic idea is the same: we obtain certain conditions for partial derivatives
at the point of a local optimum. These conditions help us either to conclude
that the local optimum already has a good value, or to prove that by relaxing
the temperature parameter we gain a certain improvement. 
We will prove the following lemma which is analogous to Lemma \ref{lem:annealing-dynamics}.

\begin{lemma}
\label{lem:diffeq}
Let $\bx(t)$ denote the local optimum found by Algorithm \ref{alg:simulated_matroidindependence}
at temperature $t < 1-1/n$ right after the ``Local search'' phase,
and let $\Phi(t) = F(\bx(t))$ denote the value of this local optimum.
Also assume that the solution found in ``Complementary solution check'' phase
of the algorithm (Steps 8-10) is always at most $\beta$.
Then the function $\Phi(t)$ satisfies
\begin{equation}
\label{eq:localgains}
  \frac{1-t}{\delta} (\Phi(t+\delta) - \Phi(t))  \geq
 (1-2\delta n^3) OPT - 2 \Phi(t) - 2 \beta t.
  \end{equation}
\end{lemma}

We proceed in two steps, again using as an intermediate bound
the notion of derivative of $F$ on the line towards the optimum:
$\G(\bx)=(\b1_C - \bx) \cdot \nabla F(\bx)$.
%= \sum_{i \in C} (1-x_i) \partdiff{F}{x_i}  - \sum_{j \notin C} x_j \partdiff{F}{x_j}.
%Similar to the work of \cite{Vondrak08} we use $\G(\bx(t))$ as in intermediate bound
%to the LHS and RHS of equation \eqref{eq:localgains}.
The plan is to relate the actual gain of the algorithm in the ``Temperature relaxation''
 phase (Steps 12-13) to $G(\bx)$, and then to argue that $G(\bx)$ can be compared
to the RHS of (\ref{eq:localgains}).
The second part relies on the submodularity of the objective function and is quite similar
to the second part of Lemma~\ref{lem:drift-bound} (although slightly more involved).

The heart of the proof is to show that by relaxing the temperature
we gain an improvement at least $\frac{\delta}{1-t} G(\bx)$. As the algorithm suggests,
the improvement in this step is related to the weight of the matching obtained in Step 12
of the algorithm. Thus the main goal is to prove that there exists a matching of weight
at least $\frac{1}{1-t} G(\bx)$. We prove this by a combinatorial argument using
the local optimality of the current fractional solution, and 
an application of K\"{o}nig's theorem on edge colorings of bipartite graphs.
We defer all details of the proof to Appendix~\ref{app:matroid}.

Finally, we arrive at a differential equation of the following form:
$$ (1-t) \Phi'(t) \geq OPT - 2 \Phi(t) - 2t\beta.$$
This differential equation is very similar to the one we obtained in 
Section~\ref{sec:unconstrained} and can be solved analytically as well.
We start from initial conditions corresponding to the $0.309$-approximation
of \cite{Vondrak09}, which implies that 
a fractional local optimum at $t_0 = \frac12 (3-\sqrt{5})$ has value
$v_0 \geq \frac12 (1-t_0) \simeq 0.309$. We prove that there is a value
$\beta > 0.325$ such that
%$$ \beta = \frac{2+\sqrt{5}}{8} \left(-5 + \sqrt{5} + \sqrt{-2 + 6 \sqrt{5}} \right) > 0.325 $$
for some value of $t$ (which turns out to be roughly $0.53$),
we get $\Phi(t) \geq \beta$. We defer details to Appendix~\ref{app:matroid}.

%%%%%%%%%%%%%%%%%%%%%%%%%%%%%%%%%%%%%%%%%%%%%%%%%%%%%%%%%%%%%%%%%%%%%%%
%%%%%%%%%%%%%%%%%%%%%%%%%%%%%%%%%%%%%%%%%%%SECTION5

\section{Hardness of approximation} 
\label{sec:hardness}

In this section, we improve the hardness of approximating several submodular maximization
problems subject to additional constraints (i.e. $\max\{f(S): S\in \mathcal{F}\}$),
assuming the value oracle model. We use the method of {\em symmetry gap}
\cite{Vondrak09} to derive these new results. 
This method can be summarized as follows.
We start with a fixed instance $\max \{f(S): S \in \cF\}$ which is symmetric
under a certain group of permutations of the ground set $X$. We consider
the multilinear relaxation of this instance, $\max \{F(\bx): \bx \in P(\cF)\}$.
We compute the {\em symmetry gap} $\gamma = \overline{OPT} / OPT$,
where $OPT = \max \{F(\bx): \bx \in P(\cF)\}$ is the optimum of the relaxed
problem and $\overline{OPT} = \max \{F(\overline{\bx}): \bx \in P(\cF) \}$ is the optimum
over all {\em symmetric} fractional solutions, i.e. satisfying $\sigma(\bar{\bx}) = \bar{\bx}$
for any $\sigma \in \cG$. Due to \cite[Theorem 1.6]{Vondrak09}, we obtain hardness
of $(1+\epsilon) \gamma$-approximation for a class of related instances, as follows.

\begin{theorem}[\cite{Vondrak09}]
\label{thm:symmetrygap}
Let $\max \{f(S): S\in \mathcal{F}\}$ be an instance of a nonnegative submodular
maximization problem with symmetry gap $\gamma=\overline{OPT}/OPT$.
Let $\mathcal{C}$ be the class of instances $\max \{\tilde{f}(S): S \in \tilde{\mathcal{F}}\}$
where $\tilde{f}$ is nonnegative submodular and $\tilde{\mathcal{F}}$ is a ``refinement``
of $\mathcal{F}$. Then for every $\eps>0$, any $(1+\epsilon)\gamma$-approximation algorithm
for the class of instances $\mathcal{C}$ would require exponentially many value queries
to $\tilde{f}(S)$. 
\end{theorem}

For a formal definition of ''refinement``, we refer to \cite[Definition 1.5]{Vondrak09}.
Intuitively, these are ''blown-up`` copies of the original family of feasible sets,
such that the constraint is of the same type as the original instance (e.g. cardinality,
matroid independence and matroid base constraints are preserved).

\paragraph{Directed hypergraph cuts.}
Our main tool in deriving these new results is a construction using a variant of the Max Di-cut
problem in {\em directed hypergraphs}. We consider the following variant of directed
hypergraphs.
% Then we use the ``symmetry gap'' machinery of Theorem \ref{thm:symmetrygap} to blow-up the instances, and create submodular functions that are hard to maximize in oracle model.

\begin{definition}
A directed hypergraph is a pair $H = (X,E)$, where $E$ is a set of directed hyperedges
$(U,v)$, where $U \subset X$ is a non-empty subset of vertices and $v \notin U$ is
a vertex in $X$. % We consider the edge as directed from $U$ to $v$.

For a set $S \subset X$, we say that a hyperedge $(U,v)$ is cut by $S$, or $(U,v) \in \delta(S)$,
if $U \cap S \neq \emptyset$ and $v \notin S$.
\end{definition}

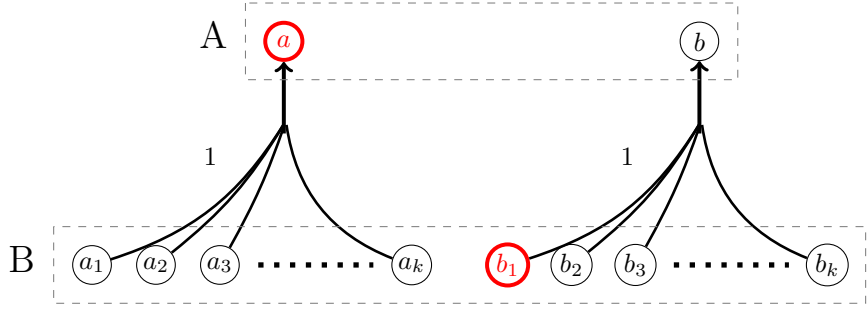
\begin{figure}
\centering
\def\sl {3}
\def \ll {6.5}
\def\hi{3.5}
\def\rl{.6}
%\begin{document}

%\begin{figure}

%\begin{tabular}{cc}
\def \Pointsize {1pt}
\begin{tikzpicture}[inner sep=1.2pt,scale=.85,pre/.style={<-,shorten <=2pt,>=stealth,thick}, post/.style={->,shorten >=1pt,>=stealth,thick}]
\tikzstyle{every node}=[draw,circle];
%\tikzstyle{every node}=[draw,shape=circle,fill=black,inner sep=\Pointsize];
\path [line width=1.5,red,inner sep=2.6pt] (0,\hi) node (a) {$a$};
\path [inner sep=2.3pt] (\ll,\hi) node (b) {$b$};
\path [outer sep=0pt] (a) +(-3,-\hi)  node (a1) {$a_1$};
\path (a) +(-2,-\hi)  node (a2) {$a_2$};
\path (a) +(-1,-\hi)  node (a3) {$a_3$};
\path (a) +(2,-\hi)  node (ak) {$a_k$};
\path [line width=1.5,color=red] (b) +(-3,-\hi)  node (b1) {$b_1$};
\path (b) +(-2,-\hi)  node (b2) {$b_2$};
\path (b) +(-1,-\hi)  node (b3) {$b_3$};
\path (b) +(2,-\hi)  node (bk) {$b_k$};

%\draw[->, >=latex, blue!20!white, line width=72pt] (4, 16) — node [black] {Optimization} +(-45:17cm);

\tikzstyle{every node}=[];
\path (a) +(0,-1.5) node (am) {} ;
\path (am) +(.03,.25) node (apm) {} ;
\path (b) +(0,-1.5 ) node (bm) {};
\path (bm) +(0.03,.25) node (bpm) {} ;

\draw [loosely dotted, line width=2pt] (a3) +(.6,0) -- +(2.4,0);
\draw [loosely dotted, line width=2pt] (b3) +(.6,0) -- +(2.4,0);

\path[->,line width=1.5pt] (am) edge (a)
	    (bm) edge  (b);
\path  [-,line width=1pt]    (a1) edge [bend right=20] node [above=15] {$1$} (apm)
	     (a2) edge [bend right=10] (apm)
	     (a3) edge [bend right=5] (apm)
	    (ak) edge [bend left] (apm)
	   (b1) edge [bend right=20] node [above=15] {$1$} (bpm)
	     (b2) edge [bend right=10] (bpm)
	     (b3) edge [bend right=5] (bpm)
	    (bk) edge [bend left]  (bpm);
%\path [-, line width=1pt]   (a) edge [] node [above] {$1-\alpha$} (b);
\draw (a) +(-\rl,\rl) node (al){};
\draw (b) +(\rl,-\rl) node (br){};
\draw (bk) + (\rl,-\rl) node (bkr){};
\draw (a1) +(-\rl,\rl) node (a1l){};
\draw [color=gray,dashed] (al) rectangle (br)  (a1l)  rectangle (bkr);

\draw (al) +(-.5,-.5) node {{\Large{A}}};
\draw (a1l) +(-.5,-.5) node {{\Large{B}}};

\end{tikzpicture} 
%\end{tabular}
%\end{figure} 

%\end{document}
\caption{Example for maximizing a submodular function subject to a matroid base constraint;
the objective function is a directed hypergraph cut function, and the constraint is
that we should pick exactly 1 element of $A$ and 1 element of $B$.}
\label{fig:hardness_matroidbase}
\end{figure}

Note that a directed hyperedge should have exactly one head. 
An example of a directed hypergraph is shown in Figure \ref{fig:hardness_matroidbase}.
We will construct our hard examples as Max Di-cut instances on directed hypergraphs.
%An input to a Max Di-cut problem is a hypergraph $H = (X,E)$,
%where $X$ is the set of vertices and $E$ is the set of hyperedges,
%and the goal is to find a cut of $H$ of maximum weight.
%A cut $(S,\overline{S})$ in $H$ cuts a directed hyperedge $(U,v)$
%if and only if $v \in \overline{S}$ and $S\cap U \neq \emptyset$.
It is easy to see that the number (or weight) of  hyperedges cut by a set $S$  is indeed submodular
 as a function of $S$.
Other types of directed hypergraphs have been considered, in particular
with hyperedges of multiple heads and tails, but a natural extension of the cut
function to such hypergraphs is no longer submodular.

In the rest of this section, 
%In subsection \ref{subsec:matroidbase}
we present our hardness result 
for maximizing submodular functions subject to a matroid base constraint.
We defer the remaining results to Appendix~\ref{app:hardness}.
%and in subsection \ref{subsec:matroidindependence}
%we improve the hardness factor of maximizing a submodular function
%subject to {\em{matroid independence (and even the cardinality) constraint}}.
%As far as we know, previous to our work the best hardness hardness factors
%for all of these problems were 1/2 \cite{FMV07}

\iffalse
\subsection{Submodular maximization over matroid bases}
\label{subsec:matroidbase}
First we consider the problem of maximizing a submodular function subject
to a matroid base constraint. Vondrak in \cite{Vondrak09} proved a hardness
factor of $(1-1/\nu)$ when the base packing number \cite[Definition 1.2]{Vondrak09}
of the matroid is $\nu<2$. Roughly speaking, the base packing number is the
maximum number of disjoint bases of the matroid that can be embedded in an instance of the problem.
On the other hand, when there are two disjoint bases (i.e. $\nu\geq 2$),
the best known hardness result is 0.5 which is the hardness of approximating any symmetric submodular function \cite{FMV07}. We show that submodular maximizing problem when the base packing number is at least 2 is  strictly harder than the unconstrained case by showing that it can not be approximated better than 0.393.
\fi

\begin{theorem}
\label{thm:hard_matroidbase}
There exist instances of the problem $\max \{f(S): S \in \cB \}$,
where $f$ is a nonnegative submodular function, $\cB$ is a collection of matroid bases
of packing number at least $2$, and any $(1-e^{-1/2}+\eps)$-approximation
for this problem would require exponentially many value queries for any $\eps>0$.
\end{theorem}

We remark that $1-e^{-1/2} < 0.394$, and only hardness of $(0.5+\eps)$-approximation
was previously known in this setting. 

\medskip
\noindent{\bf Instance 1.}
Consider the hypergraph in Figure \ref{fig:hardness_matroidbase},
with the set of vertices $X=A \cup B$ and two hyperedges $(\{a_1,\ldots,a_k\},a)$ and
$(\{b_1,\ldots,b_k\},b)$. Let $f$ be the cut function on this graph,
and let $\cM_{A,B}$ be a partition matroid whose independent sets contain at most
one vertex from each of the sets $A$ and $B$. Let $\cB_{A,B}$ be the bases of $\cM_{A,B}$
(i.e. $\cB_{A,B} = \{S: |S\cap A| = 1~\&~|S \cap B|=1\}$).
Note that there exist two disjoint bases in this matroid
and the base packing number of $\cM$ is equal to 2.  
An optimum solution is for example $S = \{a,b_1\}$ with $OPT=1$. 

In order to apply Theorem \ref{thm:symmetrygap} we need to compute the symmetry gap
of this instance $\gamma=\overline{OPT}/OPT$.
We remark in the blown-up instances, $\overline{OPT}$ corresponds to the maximum value
that any algorithm can obtain, while $OPT=1$ is the actual optimum.
The definition of $\overline{OPT}$ depends on the symmetries of our instance,
which we describe in the following lemma.

%In the next lemma we show that  the elements in the set $A(B)$ can be seen
% as if they are the same in the sense that they can be mapped to each other
% while the value of the submodular function remains unchanged. 
 
\begin{lemma}
\label{lem:symmetricgroups}
There exists a group $\cG$ of permutations such that Instance 1
is symmetric under $\cG$, in the sense that $\forall \sigma \in \cG;$
\begin{eqnarray}
\label{eq:invariant_permuation}
f(S) = f(\sigma(S)), ~~~~~~
S\in \cB_{A,B} \Leftrightarrow \sigma(S)\in \cB_{A,B}.  
\end{eqnarray}
Moreover, for any two vertices $i,j \in A$ (or $B$), the probability that
$\sigma(i) = j$ for a uniformly random $\sigma \in \cG$ is equal to $1/|A|$
(or $1/|B|$ respectively).
\end{lemma}

\begin{proof}
Let $\Pi$ be the set of the following two basic permutations
$$
\Pi=
\begin{cases} 
\sigma_1: \sigma_1(a)=b, \sigma_1(b)=a, \sigma_1(a_i)=b_i, \sigma_1(b_i)=a_i \\
\sigma_2: \sigma_2(a)=a, \sigma_2(b)=b, \sigma_2(a_i)=a_{(i \bmod k)+1}, \sigma_2(b_i)=b_i \\
%\sigma_3: \sigma_3(a)=a, \sigma_3(b)=b, \sigma_3(a_i)=a_i, \sigma_3(b_i)=b_{i+1}, 
%(b,a,b_1,b_2,\ldots,b_k,a_1,a_2,\ldots,a_k) \\
%(a,b,a_2,a_3,\ldots,a_k,a_1,b_1,b_2,\ldots,b_k) \\
%(a,b,a_1,a_2,\ldots,a_k,b_2,b_3,\ldots,b_k,b_1) \\
\end{cases}
$$
where $\sigma_1$ swaps the vertices of the two hyperedges and $\sigma_2$
only rotates the tail vertices of  one of the hyperedges.
It is easy to see that both of these permutations satisfy equation
 \eqref{eq:invariant_permuation}. Therefore, our instance is {\em{invariant}}
under each of the basic permutations and also under any permutation generated by them.
Now let $\mathcal{G}$ be the set of all the permutations that are generated by $\Pi$.
$\mathcal{G}$ is a {\em group} and under this group of symmetries all the elements
in $A$ (and $B$) are equivalent. In other words, for any three vertices $i,j,k\in A(B)$,
the number of permutations $\sigma \in G$ such that $\sigma(i) = j$ is equal to the
number of permutations such that $\sigma(i) = k$.
%$\forall i\neq j\in A(B),\ \P{\sigma\in\mathcal{G}}{\sigma(i)=j}=1/|A| (1/|B|)$) respectively
\end{proof}

Using the above lemma we may compute the {\em{symmetrization}} of a vector $\bx \in [0,1]^X$
which will be useful in computing $\overline{OPT}$ \cite{Vondrak09}.
For any vector $\bx\in [0,1]^X$, the ``symmetrization of $\bx$''  is:
\begin{equation}
\label{eq:symmetrization}
\bar{\bx}=\EE{\sigma\in\mathcal{G}}{\sigma(\bx)}=\begin{cases}
\bar{x}_a=\bar{x}_b=\frac12(x_a+x_b) \\
\bar{x}_{a_1}=\ldots=\bar{x}_{a_k}=\bar{x}_{b_1}=\ldots=\bar{x}_{b_k}
=\frac{1}{2k} \sum_{i=1}^k (x_{a_i}+x_{b_i}),
\end{cases}
\end{equation}
where $\sigma(\bx)$ denotes $\bx$ with coordinates permuted by $\sigma$.
Now we are ready to prove Theorem \ref{thm:hard_matroidbase}.\\

\begin{proof}[Theorem \ref{thm:hard_matroidbase}]
We need to compute the value of symmetry gap $\gamma=\overline{OPT}=
\max \{F(\bar{\bx}): \bx \in P(\cB_{A,B})\}$, where $F$ is the multilinear relaxation of $f$
and $P(\cB_{A,B})$ is the convex hull of the bases in $\cB_{A,B}$.
For any vector $\bx \in [0,1]^X$, we have 
\begin{equation}
\label{eq:baseproperty}
\bx\in P(\cB_{A,B}) \Leftrightarrow 
\begin{cases} x_a+x_b=1\\
\sum_{i=1}^k (x_{a_i}+x_{b_i})=1.
\end{cases}
\end{equation}
By equation \eqref{eq:symmetrization} we know that the vertices in each of the sets $A$, $B$
have the same value in $\bar{\bx}$. Using equation \eqref{eq:baseproperty},
we obtain $\bar{x}_a=\bar{x}_b=\frac12$ and $\bar{x}_{a_i}=\bar{x}_{b_i}=\frac{1}{2k}$
for all $1\leq i\leq k$,
which yields a unique symmetrized solution $\bar{\bx}=(\frac12,\frac12,
\frac{1}{2k},\ldots,\frac{1}{2k})$. 

Now we can simply compute $\overline{OPT} = F(\frac12,\frac12,\frac{1}{2k},\ldots,\frac{1}{2k})$. 
Note that by definition a hyperedge will be cut by a random set $S$
if and only if at least one of its tails are included in $S$ while its head is not included.
Therefore 
$$
\overline{OPT}=F\left(\frac12,\frac12,\frac{1}{2k},\ldots,\frac{1}{2k}\right)
 =2 \left[\frac{1}{2} \left(1-\left(1-\frac{1}{2k}\right)^k\right)\right] \simeq 1-e^{-\frac{1}{2}},
$$
for sufficiently large $k$. By applying Theorem~\ref{thm:symmetrygap}, it can be seen
that the refined instances are instances of submodular maximization over the bases
of a matroid where the ground set is partitioned into $A\cup B$ and we have to take half
of the elements of $A$ and $\frac{1}{2k}$ fraction of the elements in $B$.
Thus the base packing number of the matroid in the refined instances is also 2
which implies the theorem.
\end{proof}

\paragraph{Acknowledgment.}
We would like to thank Tim Roughgarden for stimulating discussions.

%%%%%%%%%%%%%%%%%%%%%%%%%%%%%%%%%%%%%%%%%%%%%%%%%%%%%%%%%%% APP: Miscelleneous lemmas

\appendix

\section{Miscellaneous Lemmas}
\label{app:misc}

Let $F$ be the multilinear extension of a submodular function.
The first lemma says that if we increase coordinates simultaneously, then the increase
in $F$ is {\em at most} that given by partial derivatives at the lower point,
and {\em at least} that given by partial derivatives at the upper point.

\begin{lemma}
\label{lem:submod-change}
If $F:[0,1]^X \rightarrow \RR$ is the multilinear extension of a submodular function,
and $\bx' \geq \bx$ where $\by \geq 0$, then
$$ F(\bx') \leq F(\bx) + \sum_{i \in X} (x'_i - x_i) \partdiff{F}{x_i} \Big|_\bx.$$
Similarly,
$$ F(\bx') \geq F(\bx) + \sum_{i \in X} (x'_i - x_i) \partdiff{F}{x_i} \Big|_{\bx'}.$$
\end{lemma}

\begin{proof}
Since $F$ is the multilinear extension of a submodular function,
we know that $\mixdiff{F}{x_i}{x_j} \leq 0$ for all $i,j$ \cite{CCPV07}.
This means that whenever $\bx \leq \bx'$, the partial derivatives
at $\bx'$ cannot be larger than at $\bx$:
$$ \partdiff{F}{x_i} \Big|_{\bx} \geq \partdiff{F}{x_i} \Big|_{\bx'}. $$
Therefore, between $\bx$ and $\bx'$, the highest partial derivatives
are attained at $\bx$, and the lowest at $\bx'$.
By integrating along the line segment between $\bx$ and $\bx'$, we obtain
$$ F(\bx') - F(\bx) = \int_0^1 (\bx'-\bx) \cdot \nabla F(\bx + t (\bx'-\bx)) dt
 = \sum_{i \in X} \int_0^1 (x'_i-x_i) \partdiff{F}{x_i} \Big|_{\bx + t (\bx'-\bx)} dt.$$
If we evaluate the partial derivatives at $\bx$ instead, we get
$$ F(\bx') - F(\bx) \leq \sum_{i \in X} (x'_i - x_i) \partdiff{F}{x_i} \Big|_\bx.$$
If we evaluate the partial derivatives at $\bx'$, we get
$$ F(\bx') - F(\bx) \geq \sum_{i \in X} (x'_i - x_i) \partdiff{F}{x_i} \Big|_{\bx'}.$$
\end{proof}

For a small increase in each coordinate, the partial derivatives give
a good approximation of the change in $F$; this is a standard analytic argument,
which we formalize in the next lemma.

\begin{lemma}
\label{lem:Taylor}
Let $F:[0,1]^X \rightarrow \RR$ be twice differentiable,
$\bx \in [0,1]^X$ and $\by \in [-\delta,\delta]^X$. Then
$$ \Big|F(\bx+\by) - F(\bx) - \sum_{i \in X} y_i \partdiff{F}{x_i} \Big|_\bx \Big|
 \leq \delta^2 n^2 \ \sup \Big| \mixdiff{F}{x_i}{x_j} \Big|,$$
 where the supremum is taken over all $i,j$ and all points in $[0,1]^X$.
\end{lemma}

\begin{proof}
Let $M = \sup |\mixdiff{F}{x_i}{x_j}|$.
Since $F$ is twice differentiable, any partial derivative can change by at most
$\delta M$ when a coordinate changes by at most $\delta$. Hence,
$$ -\delta n M \leq \partdiff{F}{x_i} \Big|_{\bx+t\by} - \partdiff{F}{x_i} \Big|_\bx
 \leq \delta n M$$
for any $t \in [0,1]$. By the fundamental theorem of calculus,
$$ F(\bx+\by) = F(\bx) + \sum_{i \in X} \int_0^1 y_i \partdiff{F}{x_i} \Big|_{\bx+t\by} dt 
 \leq F(\bx) + \sum_{i \in X} y_i \left( \partdiff{F}{x_i} \Big|_\bx + \delta n M \right)
  \leq F(\bx) + \sum_{i \in X} y_i \partdiff{F}{x_i} \Big|_\bx + \delta^2 n^2 M.$$
Similarly we get $F(\bx+\by) \geq F(\bx) + \sum_{i \in X} y_i \partdiff{F}{x_i} \Big|_\bx
  - \delta^2 n^2 M.$
\end{proof}

The following ``threshold lemma`` appears as Lemma~A.4 in \cite{Vondrak09}.
We remark that the expression $\E{f(T_{>\lambda}(\bx))}$ defined below
is an alternative definition of the Lov\'asz extension of $f$.

\begin{lemma}[Threshold Lemma]
\label{lem:threshold}
For $\by \in [0,1]^X$ and $\lambda \in [0,1]$, define $T_{>\lambda}(\by)
 = \{ i: y_i > \lambda\}$. If $F$ is the multilinear extension of a submodular
function $f$, then for $\lambda \in [0,1]$ uniformly random
$$ F(\by) \geq \E{f(T_{>\lambda}(\by))}.$$
\end{lemma}

Since we apply this lemma in various places of the paper let us describe
some applications of it in detail.

\begin{example}
In this example we apply the threshold lemma to the vector $\bx=p\b1_{A\cap C}+(1-p)\b1_{B\cap C}$.
Here $C$ represents the optimum set, $B=\overline{A}$  and $1/2<p<1$.
If $\lambda\in[0,1]$ is chosen uniformly at random  we know $0 < \lambda \leq 1-p$
with probability $1-p$, $1-p <\lambda \leq p$ with probability $2p-1$
and $p<\lambda \leq 1$ with probability $1-p$. Therefore by Lemma~\ref{lem:threshold} we have:
\begin{eqnarray*}
F(\bx)  & \geq & (1-p) \E{f(T_{>\lambda}(\bx)) | \lambda \leq 1-p} +
 (2p-1) \E{f(T_{>\lambda}(\bx)) | 1-p<\lambda \leq p}+
 (1-p) \E{f(T_{>\lambda}(\bx)) | p<\lambda \leq 1}\\
&=& (1-p) \E{f(C)} + (2p-1) \E{f(A\cap C)}+(1-p) \E{f(\emptyset)}\\
\end{eqnarray*}
or equivalently we can write
\begin{equation}
\label{eq:threshold_easy}
\T{p}{0}{1-p}{0} \geq  (1-p)~\T{1}{0}{1}{0} + (2p-1)~\T{1}{0}{0}{0}+(1-p)~\T{0}{0}{0}{0}.
\end{equation}
\end{example}

In the next example we consider a more complicated application of the threshold lemma.
\begin{example}
Consider the vector  $\bx$ where $x_i=1$ for  $i\in C$, $x_i=t$ for $i\in A\setminus C$ and $x_i<t$ for $i\in B\setminus C$. In this case, we denote
$$ F(\bx) = \T{1}{t}{1}{x}.$$
Again $C$ is the optimal set and $B=\overline{A}$. In this case if we apply the threshold lemma,
we get a random set which can contain a part of the block $B \setminus C$.
In particular, observe that if $\lambda \leq t$, then $T_{>\lambda}(\bx)$ contains all the
elements in $\overline{B\setminus C}$, and depending on the value of $\lambda$,  elements in $B\setminus C$ that are greater than $\lambda$. We denote the value of such a set by 
$$ f(T_{>\lambda}(\bx)) = \TTR{1}{1}{1}{1}{0} $$
where the right-hand lower block is divided into two parts depending on the threshold $\lambda$.
%we have $\lambda>t$ with probability $1-t$ and $\lambda<t$ with probability $t$, but if  
Therefore 
\begin{equation*}
F(\bx)\geq t~ \E{f(T_{>\lambda}(\bx)) | \lambda \leq t}
 + (1-t)\E{f(T_{>\lambda}(\bx)) | \lambda > t },
\end{equation*}
can be written equivalently as
\begin{equation}
\label{eq:threshold_hard}
\T{1}{t}{1}{x} \geq  t~\E{\, \TTR{1}{1}{1}{1}{0} \, \Big| \, \lambda \leq t}
  +(1-t)~\T{1}{0}{1}{0}.
\end{equation}
%Note that we used the notation $\TTR{1}{1}{1}{1}{0}$ as a representation of %$\E{f(T_{>\lambda}%(\bx) | t <\lambda }$, the bottom left cell of the diagram is divided into %two cells to express that condition on the value of $\lambda<t$, the elements in $B\setminus C$ %will be decomposed into two sets.
\end{example}

A further generalization of the threshold lemma is the following, which is also useful
in our analysis. (See \cite[Lemma A.5]{Vondrak09}.)

\begin{lemma}
\label{lem:threshold2}
For any partition $X = X_1 \cup X_2$,
$$ F(\bx) \geq \E{f((T_{>\lambda_1}(\bx) \cap X_1) \cup (T_{>\lambda_2}(\bx) \cap X_2))} $$
where $\lambda_1,\lambda_2$ are independent and uniformly random in $[0,1]$.
\end{lemma}

%%%%%%%%%%%%%%%%%%%%%%%%%%%%%%%%%%%%%%%%%%
%%%%%%%%%%%%%%%%%%%%%%%%%%%%%%%%%%%%%%%APP: Unconstrained submodular maximization

\section{Analysis of the $0.41$-approximation}
\label{app:unconstrained}

Here we finish the analysis of the simulated annealing algorithm
for unconstrained submodular maximization (Theorem~\ref{thm:0.41-approx}).
Consider Lemma~\ref{lem:annealing-dynamics} in the limit when $\delta \rightarrow 0$.
It gives the following differential inequality:
\begin{equation}
\label{eq:diff-eq}
(1-p) \Phi'(p) \geq OPT - 2 \Phi(p) - (2p-1) \beta.
\end{equation}

We assume here that $\delta$ is so small that the difference
between the solution of this differential inequality and the actual
behavior of our algorithm is negligible. 
(We could replace $OPT$ by $(1-\epsilon) OPT$, carry out the analysis
and then let $\epsilon \rightarrow 0$; however, we shall spare the reader of this annoyance.)
Our next step is to solve this differential equation, given certain initial conditions.
Without loss of generality, we assume that $OPT=1$.

\begin{lemma}
\label{lem:diff-solution}
Assume that $OPT=1$.
Let $\Phi(p)$ denote the value of the solution at temperature $t=1-p$.
Assume that $\Phi(p_0) = v_0$ for some $p_0 \in (\frac12,1)$, and $f(\overline{A(p)}) \leq \beta$
for all $p$. Then for any $p \in (p_0,1)$,
$$ \Phi(p) \geq \frac12 (1-\beta) + 2 \beta (1-p) -
 \frac{(1-p)^2}{(1-p_0)^2} \left(\frac12 (1-\beta) + 2 \beta (1-p_0) - v_0 \right).$$ 
\end{lemma}

\begin{proof}
We rewrite Equation~(\ref{eq:diff-eq}) using the following trick:
$$ (1-p)^3 \frac{d}{dp}((1-p)^{-2} \Phi(p)) = (1-p)^3 (2 (1-p)^{-3} \Phi(p) + (1-p)^{-2} \Phi'(p))
 = 2 \Phi(p) + (1-p) \Phi'(p).$$
Therefore, Lemma~\ref{lem:annealing-dynamics} states that
$$ (1-p)^3 \frac{d}{dp}(p^{-2} \Phi(p)) \geq OPT - (2p-1) \beta 
= 1 - (2p-1) \beta = 1-\beta + 2 \beta (1-p) $$
which is equivalent to
$$ \frac{d}{dp}((1-p)^{-2} \Phi(p)) \geq \frac{1-\beta}{(1-p)^3} + \frac{2 \beta}{(1-p)^2}.$$
For any $p \in (p_0,1)$, the fundamental theorem of calculus implies that
\begin{eqnarray*}
(1-p)^{-2} \Phi(p) - (1-p_0)^{-2} \Phi(p_0)
 & \geq & \int_{p_0}^{p} \left( \frac{1-\beta}{(1-\tau)^3} + \frac{2 \beta}{(1-\tau)^2} \right) d\tau
 \\
& = & \left[ \frac{1-\beta}{2 (1-\tau)^2} + \frac{2 \beta}{1-\tau} \right]_{p_0}^p \\
& = & \frac{1-\beta}{2(1-p)^2} + \frac{2 \beta}{1-p} - \frac{1-\beta}{2(1-p_0)^2} - \frac{2 \beta}{1-p_0}.
\end{eqnarray*}
Multiplying by $(1-p)^2$, we obtain
$$ \Phi(p) \geq \frac12 (1-\beta) + 2 \beta (1-p) +
 \frac{(1-p)^2}{(1-p_0)^2} \left(\Phi(p_0) - \frac12(1-\beta) - 2 \beta (1-p_0) \right).$$
\end{proof}

In order to use this lemma,
recall that the parameter $\beta$ is an upper bound on the values of $f(\overline{A})$ throughout
the algorithm. This means that we can choose $\beta$ to be our "target value":
if $f(\overline{A})$ achieves value more than $\beta$ at some point, we are done.
If $f(\overline{A})$ is always upper-bounded by $\beta$, we can use Lemma~\ref{lem:diff-solution},
hopefully concluding that for some $p$ we must have $\Phi(p) \geq \beta$.

In addition, we need to choose a suitable initial condition.
As a first attempt, we can try to plug in $p_0 = 1/2$ and $v_0 = 1/4$ as a starting point
(the uniformly random $1/4$-approximation provided by \cite{FMV07}). We would obtain
$$ \Phi(p) \geq \frac12 (1-\beta) + 2 \beta (1-p) - (1+2\beta) (1-p)^2.$$
However, this is not good enough. For example, if we choose $\beta = 2/5$ as our target value,
we obtain $ \Phi(p) \geq \frac{3}{10} + \frac{4}{5} (1-p) - \frac{9}{5} (1-p)^2.$
In can be verified that this function stays strictly below $2/5$ for all $p \in [\frac12,1]$.
So this does not even match the performance of the $2/5$-approximation of \cite{FMV07}.

As a second attempt, we can use the $2/5$-approximation itself as a starting point.
The analysis of \cite{FMV07} implies that if $A$ is a local optimum for $p_0=2/3$,
we have either $f(\overline{A}) \geq 2/5$, or $F(\bx_p(A)) \geq 2/5$. This means that we can use
the starting point $p_0 = 2/3, v_0 = 2/5$ with a target value of $\beta = 2/5$
(effectively ignoring the behavior of the algorithm for $p < 2/3$).
Lemma~\ref{lem:diff-solution} gives
$$ \Phi(p) \geq \frac{3}{10} + \frac{4}{5} (1-p) - \frac{3}{2} (1-p)^2.$$
The maximum of this function is attained at $p_0 = 11/15$ which gives $\Phi(p_0) \geq 61/150 > 2/5$.
This is a good sign - however, it does not imply that the algorithm actually achieves
a $61/150$-approximation, because we have used $\beta = 2/5$ as our target value.
(Also, note that $61/150 < 0.41$, so this is not the way we achieve our main result.)

In order to get an approximation guarantee better than $2/5$, we need to revisit the analysis of \cite{FMV07}
and compute the approximation factor of a local optimum as a function of the temperature $t=1-p$
and the complementary solution $f(\overline{A}) = \beta$.

\begin{lemma}
\label{lem:starting-point}
Assume $OPT = 1$.
Let $q \in [\frac{1}{3},\frac{1}{1+\sqrt{2}}]$, $p=1-q$ and
let $A$ be a local optimum with respect to $F(\bx_p(A))$.
Let $\beta = f(\overline{A})$. Then 
$$ F(\bx_p(A)) \geq \frac12 (1-q^2) - q(1-2q) \beta.$$
\end{lemma}

\begin{proof}
$A$ is a local optimum with respect to the objective function $F(\bx_p(A))$.
We denote $\bx_p(A)$ simply by $\bx$. Let $C$ be a global optimum and $B = \overline{A}$.
As we argued in the proof of Lemma~\ref{lem:drift-bound}, we have
$$ \T{p}{p}{q}{q} \geq \T{p}{0}{q}{q} $$
and also
$$ \T{p}{p}{q}{q} \geq \T{p}{p}{1}{q} $$
We apply Lemma~\ref{lem:threshold2} which states that
$F(\bx) \geq \E{f((T_{>\lambda_1}(\bx) \cap C) \cup (T_{>\lambda_2}(\bx) \setminus C))}$,
where $\lambda_1, \lambda_2$ are independent and uniformly random in $[0,1]$.
This yields the following (after dropping some terms which are nonnegative):
\begin{eqnarray}
\label{eq:unconst_intial_localopt1}
 \T{p}{p}{q}{q} \geq \T{p}{p}{1}{q} \geq
  pq~\T{1}{0}{1}{0} + p(p-q)~\T{1}{0}{0}{0} + q^2~\T{1}{0}{1}{1} + (p-q)q~\T{1}{0}{0}{1} \\
  \label{eq:unconst_intial_localopt2}
 \T{p}{p}{q}{q} \geq \T{p}{0}{q}{q}
\geq pq~\T{1}{0}{1}{0} + p(p-q)~\T{1}{1}{1}{0} + q^2~\T{0}{0}{1}{0} + (p-q)q~\T{0}{1}{1}{0} 
\end{eqnarray}
The first term in each bound is $pq \cdot OPT$.
However, to make use of the remaining terms, we must add some terms on both sides.
The terms we add are $\frac12 (-p^3 + p^2 q + 2pq^2) f(A) + \frac12 (p^3 + p^2 q -2pq^2 -2q^3) f(B)$;
it can be verified that both coefficients are nonnegative for $q \in [\frac13, \frac{1}{1+\sqrt{2}}]$.
Also, the coefficients are chosen so that they sum up to $p^2 q - q^3 = q (p^2 - q^2) = q (p-q)$,
the coefficient in front of the last term in each equation. Using submodularity, we get
\begin{eqnarray}
 & \frac12 (-p^3 + p^2 q + 2pq^2) \T{1}{1}{0}{0} + \frac12 (p^3 + p^2 q -2pq^2 -2q^3) \T{0}{0}{1}{1} + (p-q)q~\T{1}{0}{0}{1} \nonumber \\
 = & \frac12 (-p^3 + p^2 q + 2pq^2) \left[ \T{1}{1}{0}{0} + \T{1}{0}{0}{1} \right]  + 
 \frac12 (p^3 + p^2 q -2pq^2 -2q^3) \left[ \T{0}{0}{1}{1} + \T{1}{0}{0}{1} \right] \nonumber \\
\label{eq:unconst_intial_localopt3}
 \geq & \frac12 (-p^3 + p^2 q + 2pq^2) \T{1}{0}{0}{0} + 
 \frac12 (p^3 + p^2 q -2pq^2 -2q^3) \T{1}{0}{1}{1}.
\end{eqnarray}
Similarly, we get
\begin{eqnarray}
 & \frac12 (-p^3 + p^2 q + 2pq^2) \T{1}{1}{0}{0} + \frac12 (p^3 + p^2 q -2pq^2 -2q^3) \T{0}{0}{1}{1} + (p-q)q~\T{0}{1}{1}{0} \nonumber \\
 = & \frac12 (-p^3 + p^2 q + 2pq^2) \left[ \T{1}{1}{0}{0} + \T{0}{1}{1}{0} \right]
  + \frac12 (p^3 + p^2 q -2pq^2 -2q^3) \left[ \T{0}{0}{1}{1} + \T{0}{1}{1}{0} \right] \nonumber \\
  \label{eq:unconst_intial_localopt4}
 \geq & \frac12 (-p^3 + p^2 q + 2pq^2) \T{1}{1}{1}{0} +
  \frac12 (p^3 + p^2 q -2pq^2 -2q^3) \T{0}{0}{1}{0}.
\end{eqnarray}
Putting equations \eqref{eq:unconst_intial_localopt1}, \eqref{eq:unconst_intial_localopt2} \eqref{eq:unconst_intial_localopt3} and \eqref{eq:unconst_intial_localopt4} all together, we get
\begin{eqnarray*}
& 2~\T{p}{p}{q}{q} + (-p^3 + p^2 q + 2pq^2) \T{1}{1}{0}{0}
 + (p^3 + p^2 q -2pq^2 -2q^3) \T{0}{0}{1}{1} \\
\geq & 2pq~\T{1}{0}{1}{0} + (p(p-q) + \frac12 (-p^3 + p^2 q + 2pq^2))
 \left[ \T{1}{1}{1}{0} + \T{1}{0}{0}{0} \right] \\
 & + (q^2 + \frac12 (p^3 + p^2 q - 2pq^2 - 2q^3)) 
 \left[ \T{1}{0}{1}{1} + \T{0}{0}{1}{0} \right]\\
 = & 2pq~\T{1}{0}{1}{0} + \frac12 p^2 \left[ \T{1}{1}{1}{0} + \T{1}{0}{0}{0}
  + \T{1}{0}{1}{1} + \T{0}{0}{1}{0} \right].
 \end{eqnarray*}
where the simplification came about by using the elementary relations
 $p(p-q) = p(p-q)(p+q) = p(p^2-q^2)$ and $q^2 = q^2 (p+q)$.
Submodularity implies 
$$ \T{1}{0}{0}{0} + \T{0}{0}{1}{0} \geq \T{1}{0}{1}{0} = OPT $$
and 
$$ \T{1}{1}{1}{0} + \T{1}{0}{1}{1} \geq \T{1}{0}{1}{0} = OPT, $$
so we get, replacing the respective diagrams by $F(\bx)$, $f(A)$ and $f(B)$,
$$ 2 F(\bx) + (-p^3 + p^2 q + 2pq^2) f(A) + (p^3 + p^2 q -2pq^2 -2q^3) f(B)
 \geq (2pq + p^2) OPT = (1-q^2) OPT $$
again using $(p+q)^2 = 1$.
Finally, we assume that $f(A) \leq \beta$ and $f(B) \leq \beta$, which means
$$ 2 F(\bx) \geq (1-q^2) OPT - (2p^2 q - 2q^3) \beta = (1-q^2) OPT - 2q (p-q) \beta
 = (1-q^2) OPT - 2q (1-2q) \beta.$$
\end{proof}

Now we can finally prove Theorem~\ref{thm:0.41-approx}. Consider Lemma~\ref{lem:diff-solution}.
Starting from $\Phi(p_0) = v_0$, we obtain the following bound for any $p \in (p_0,1)$:
$$ \Phi(p) \geq \frac12 (1-\beta) + 2 \beta (1-p) -
 \frac{(1-p)^2}{(1-p_0)^2} \left(\frac12 (1-\beta) + 2 \beta (1-p_0) - v_0 \right).$$
By optimizing this quadratic function, we obtain that the maximum is attained at
$p_1 = \frac{\beta (1-p_0)^2}{(1-\beta)/2 + 2 \beta (1-p_0) - v_0}$ and the corresponding bound is
$$ \Phi(p_1) \geq \frac12 (1-\beta) + \frac{\beta^2 (1-p_0)^2}{(1-\beta)/2 + 2\beta (1-p_0) - v_0}.$$
Lemma~\ref{lem:starting-point} implies that a local optimum at temperature $q=1-p_0
 \in [\frac13, \frac{1}{1+\sqrt{2}}]$ has value
$ v_0 \geq \frac12 (1-q^2) - q(1-2q) \beta = p_0 - \frac12 p_0^2 - (1-p_0)(2 p_0 -1) \beta.$ 
Therefore, we obtain
$$ \Phi(p_1) \geq \frac12 (1-\beta) + \frac{\beta^2 (1-p_0)^2}{(1-\beta)/2
 + 2\beta (1-p_0) - p_0 + \frac12 p_0^2 + (1-p_0)(2 p_0 -1) \beta}.$$ 
We choose $p_0 = \sqrt{2} / (1+\sqrt{2})$ and solve for a value of $\beta$ such that
$\Phi(p_1) \geq \beta$. This value can be found as a solution of a quadratic equation
and is equal to
$$ \beta = \frac{1}{401} \left(37 + 22\sqrt{2} + (30\sqrt{2}+14) \sqrt{-5\sqrt{2} + 10} \right).$$
It can be verified that $\beta > 0.41$.
This completes the proof of Theorem~\ref{thm:0.41-approx}.

%%%%%%%%%%%%%%%%%%%%%%%%%%%%%%%%%%%%%
%%%%%%%%%%%%%%%%%%%%%%%%%%%%%%APP: Tight example

\section{Upper Bounding the Performance of the Simulated Annealing Algorithm}
\label{app:tightexample}

In this section we show that the simulated annealing algorithm
\ref{alg:simulated_unconstrained} for unconstrained submodular maximization,
does not give a half approximation even on instances of the directed 
maximum cut problem. We provide a directed graph $G$ (found by an LP-solver)
and a set of local optimums for all values of $p\in [1/2,1]$, such the value
of $f$ on each of them or their complement is at most $0.486$ of $OPT$. 

\begin{theorem}
\label{thm:tightexample}
There exists an instance of the unconstrained submodular maximization problem,
such that the approximation factor of Algorithm~\ref{alg:simulated_unconstrained}
is $17/35 < 0.486$.
\end{theorem}

%We remark that there is some ambiguity in how the algorithm resolves ties
%in searching for a local optimum. We assume here that the algorithm is ''unlucky``
%in that it resolves ties in a way that leads to a factor of $17/35$.
%However, the instance can be perturbed so that no ties occur and the algorithm
%is forced to follow this path of computation.

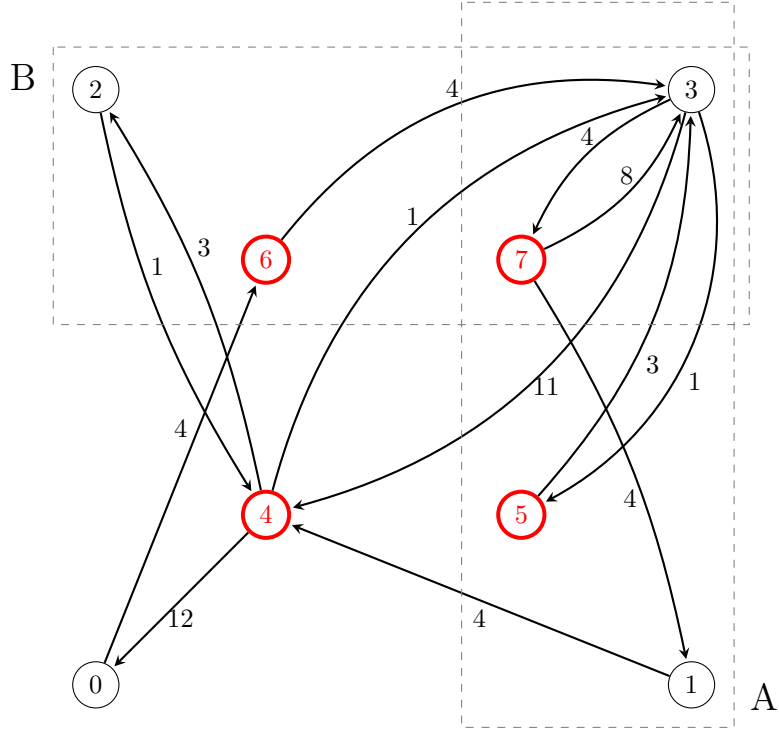
\begin{figure}
\centering
\def \Pointsize {1.4pt}
\def\sl {3}
\def \ll {7}
%\begin{document}

%\begin{figure}

\def \Pointsize {1pt}
\begin{tikzpicture}[scale=.8,pre/.style={<-,shorten <=2pt,>=stealth,thick}, post/.style={->,shorten >=1pt,>=stealth,thick}]
\tikzstyle{every node}=[draw,circle];
\path[line width=1.5pt,color=red] (45:\sl cm) node (v7) {$7$};
\path[line width=1.5pt,color=red] (135:\sl cm) node (v6) {$6$};
\path[line width=1.5pt,color=red] (225:\sl cm) node (v4) {$4$};
\path[line width=1.5pt,color=red] (315:\sl cm) node (v5) {$5$};
\path (45:\ll cm) node (v3) {$3$};
\path (135:\ll cm) node (v2) {$2$};
\path (225:\ll cm) node (v0) {$0$};
\path (315:\ll cm) node (v1) {$1$};

%\draw[->, >=latex, blue!20!white, line width=72pt] (4, 16) — node [black] {Optimization} +(-45:17cm);

\tikzstyle{every node}=[];

\path[->] (v7) edge [post,bend right=20] node [above] {8} (v3)
	      (v3) edge [post,bend right=20] node [above] {4} (v7)
	      (v4) edge [post,bend left=30] node [above] {1} (v3)
	      (v3) edge [post,bend left=30] node [below] {11} (v4)
	      (v5) edge [post,bend right=20] node [below=10] {3} (v3)
	      (v3) edge [post, bend left=40] node [below=10] {1} (v5)
	      (v6) edge [post, bend left] node [above] {4} (v3) 
	      (v0) edge [post] node [above=10] {4} (v6) 
	      (v4) edge [post] node [below] {12} (v0) 
	      (v4) edge [post, bend right=10] node [above=10] {3} (v2) 
	      (v2) edge [post, bend right=10] node [above=10] {1} (v4) 
	      (v7) edge [post, bend left=10] node [above=-20] {4} (v1) 
	      (v1) edge [post] node [below] {4} (v4) ;

\draw [color=gray,dashed] (135:8cm) rectangle (10:6cm)  (80:6.5cm) rectangle (-45:8cm);
\draw (135:8cm) +(-.5,-.5) node {{\Large{B}}};
\draw (-45:8cm) +(.5,.5) node {{\Large{A}}};

\end{tikzpicture}

%\end{figure}
%\end{document}
%
\caption{Hard instance of the unconstrained submodular maximization problem,
 where Algorithm \ref{alg:simulated_unconstrained} may get value no more than $17$.
 The bold vertices $\{4,5,6,7\}$ represents the optimum set with value $OPT=35$.}
\label{fig:tightexample}
\end{figure}

\begin{proof}
Let $f$ be the cut function of the directed graph $G$ in Figure \ref{fig:tightexample}.
We show that the set $A=\{1,3,5,7\}$ is a local optimum for all $p\in [\frac{1}{2},
\frac{3}{4}]$ and the set $B=\{2,4,6,8\}$ is a local optimum for all $p\in [\frac{3}{4},1]$.
Moreover, since we have $F(\bx_{3/4}(A)) = F(\bx_{3/4}(B)=16.25$, it is possible
that in a run of the simulated annealing algorithm \ref{alg:simulated_unconstrained},
the set $A$ is chosen and remains as a local optimum fort $p=1/2$ to $p=3/4$.
Then the local optimum changes to $B$ and remains until the end of the algorithm. 
If the algorithm follows this path then its approximation ratio
is $17/35$. This is because the value of the optimum set
$f(\{4,5,6,7\})=35$, while $\max\{f(A),f(B),f(\overline{A}),f(\overline{B})\} = 17$.
We remark that even sampling from $A,\overline{A}$ (or from $B,\overline{B}$)
with probabilities $p,q$ does not give value more than 17.

It remains to show that the set $A$ is in fact a local optimum for all
$p\in[\frac{1}{2},\frac{3}{4}]$. We just need to show that all the elements
in $A$ have a  non-negative partial derivative and the elements in $\overline{A}$
have a non-positive partial derivative. Let $p\in[\frac{1}{2}, \frac{3}{4}]$
and $q=1-p$, then:

\begin{center}
\begin{tabular}{lcl}
$\partdiff{F}{x_0} = -12q+4p \leq 0$ & & $\partdiff{F}{x_1} =  4p  - 4(1-q) = 0$ \\
$\partdiff{F}{x_2} =  -3q +p \leq 0$ & & $\partdiff{F}{x_3} = -11p+5q+11p-5q=0$\\
 $\partdiff{F}{x_4} =  15p-q-15p+q = 0$ & & $\partdiff{F}{x_5} = -p+3q \geq 0$ \\
  $\partdiff{F}{x_6} =  -4q + 4q = 0$ & & $\partdiff{F}{x_7} = -4p+12q\geq 0$ 
%\end{eqnarray*} 
\end{tabular}
\end{center}

Therefore, $A$ is a local optimum for $p\in[\frac12,\frac34]$.
Similarly, it can be shown that $B$ is a local optimum for $p\in[\frac{3}{4},1]$
which concludes the proof.
\end{proof}

%%%%%%%%%%%%%%%%%%%%%%%%%%%%%%%%%%%%%%%%%%
%%%%%%%%%%%%%%%%%%%%%%%%%%%%%%%%%%%APP: Matroid independence

\section{Analysis of the $0.325$-approximation}
\label{app:matroid}

Our first goal is to prove Lemma~\ref{lem:diffeq}. As we discussed,
the key step is to compare the gain in the temperature relaxation step
to the value of the derivative on the line towards the optimum,
$G(\bx) = (\b1_C - \bx) \cdot \nabla F(\bx)$.
We prove the following.

\begin{lemma}
\label{lem:temp-gain}
Let $\bx(t)$ be the the local optimum at time $t<1-1/n$. Then
$$ \frac{1-t}{\delta} (F(\bx(t+\delta)) - F(\bx(t))) \geq 
  \G(\bx(t)) - n^2 \delta \, \sup \Big| \mixdiff{F}{x_i}{x_j} \Big|.$$
\end{lemma}

This lemma can be compared to the first part of the proof of Lemma \ref{lem:drift-bound},
which is not very complicated in the unconstrained case. As we said, the main difficulty here
is that relaxing the temperature does not automatically allow us to increase
all the coordinates with a positive partial derivative.
The reason is that the new fractional solution might not belong to $P_{t+\delta}(\cM)$.
Instead, the algorithm modifies coordinates according to a certain maximum-weight matching
found in Step 12. The next lemma shows that the weight of this matching
is comparable to $\G(\bx)$.

\begin{lemma}
\label{lem:matching}
Let $\bx = \frac{1}{N} \sum_{\ell=1}^{N} \b1_{I_\ell} \in P_t(\cM)$
be a fractional local optimum, and $C \in \cI$ a global optimum.
Assume that $(1-t)N \geq n$. Let $G_x$ be the fractional exchange graph
defined in Def.~\ref{def:ex-graph}.
Then $G_x$ has a matching $M$ of weight
$$ w(M) \geq \frac{1}{1-t} \G(\bx).$$
\end{lemma}

\begin{proof}
We use a basic property of matroids (see \cite{Schrijver}) which says
that for any two independent sets $C, I \in \cI$, there is a
mapping $m:C \setminus I \rightarrow (I \setminus C) \cup \{0\}$ such that
for each $i \in C \setminus I$, $I - m(i) + i$ is independent,
and each element of $I \setminus C$ appears at most once as $m(i)$.
I.e., $m$ is a matching, except for the special element $0$ which can be used
as $m(i)$ whenever $I+i \in \cI$.
Let us fix such a mapping for each pair $C,I_\ell$, and denote the respective
mapping by $m_\ell:C \setminus I_\ell \rightarrow I_\ell \setminus C$.

Denote by $W$ the sum of all positive edge weights in $G_x$. We estimate
$W$ as follows. For each $i \in A \cap C$ and each edge $(i,\ell)$, we have
$i \in A \cap C \setminus I_\ell$ and by Lemma~\ref{lem:best-match}
$$ w_{i\ell} = \partdiff{F}{x_i} - \partdiff{F}{x_{b_\ell(i)}}
 \geq \partdiff{F}{x_i} - \partdiff{F}{x_{m_\ell(i)}}. $$
Observe that for $i \in (C \setminus A) \setminus I_\ell$, we get
$$ 0 \geq \partdiff{F}{x_i} - \partdiff{F}{x_{m_\ell(i)}} $$
because otherwise we could replace $I_\ell$ by $I_\ell-m_\ell(i)+i$,
which would increase the objective function (and for elements outside of $A$,
we have $x_i < t$, so $x_i$ can be increased).
Let us add up the first inequality over all elements $i \in A \cap C \setminus I_\ell$
and the second inequality over all elements $i \in (C \setminus A) \setminus I_\ell$:
$$ \sum_{i \in A \cap C \setminus I_\ell} w_{i\ell} \geq
 \sum_{i \in C \setminus I_\ell} \left( \partdiff{F}{x_i}
  - \partdiff{F}{x_{m_\ell(i)}} \right) 
  \geq \sum_{i \in C \setminus I_\ell} \partdiff{F}{x_i}
   - \sum_{j \in I_\ell \setminus C} \partdiff{F}{x_j}  $$
where we used the fact that each element of $I_\ell \setminus C$
appears at most once as $m_\ell(i)$, and $\partdiff{F}{x_j} \geq 0$
for any element $j \in I_\ell$ (otherwise we could remove it
and improve the objective value).
Now it remains to add up these inequalities over all $\ell=1,\ldots,N$:
$$ \sum_{\ell=1}^{N} \sum_{i \in A \cap C \setminus I_\ell} w_{i \ell}
 \geq \sum_{\ell=1}^{N} \left(  \sum_{i \in C \setminus I_\ell} \partdiff{F}{x_i}
   - \sum_{j \in I_\ell \setminus C} \partdiff{F}{x_j} \right)
    = N \sum_{i \in C} (1-x_i) \partdiff{F}{x_i}
    - N \sum_{j \notin C} x_j \partdiff{F}{x_j} $$
using $x_i = \sum_{\ell: i \in I_\ell} \frac{1}{N}$.
The left-hand side is a sum of weights over a subset of edges.
Hence, the sum of all positive edge weights also satisfies
$$ W \geq N \sum_{i \in C} (1-x_i) \partdiff{F}{x_i}
    - N \sum_{j \notin C} x_j \partdiff{F}{x_j} = N \cdot \G(\bx). $$
Finally, we apply K\"{o}nig's theorem on edge colorings of bipartite graphs:
Every bipartite graph of maximum degree $\Delta$ has an edge coloring using
at most $\Delta$ colors.
The degree of each node $i \in A$ is the number of sets $I_\ell$ not containing $i$,
which is $(1-t)N$, and the degree of each node $\ell \in [N]$ is at most the number
of elements $n$, by assumption $n \leq (1-t)N$. By K\"{o}nig's theorem,
there is an edge coloring using $(1-t)N$ colors. Each color class is a matching,
and by averaging, the positive edge weights in some color class have total weight
$$ w(M) \geq \frac{W}{(1-t)N} \geq \frac{1}{1-t} \G(\bx).$$
\end{proof}

The weight of the matching found by the algorithm corresponds to how
much we gain by increasing the parameter $t$. Now we can prove Lemma \ref{lem:temp-gain}.

\medskip

\begin{proof}[Lemma \ref{lem:temp-gain}]
Assume the algorithm finds a matching $M$. By Lemma~\ref{lem:matching},
its weight is
$$ w(M) = \sum_{(i,\ell) \in M} \left(\partdiff{F}{x_i} - \partdiff{F}{x_{b_\ell(i)}} \right)
 \geq \frac{1}{1-t} \G(\bx(t)).
 $$
 %\left(\sum_{i \in C} (1-x_i) \partdiff{F}{x_i}
 %- \sum_{j \notin C} x_j \partdiff{F}{x_j} \right).
 %$$
If we denote by $\tilde{\bx}(t)$ the fractional solution right after the ``Temperature relaxation'' phase, we have
$$ \tilde{\bx}(t) = \bx(t) + \delta \sum_{(i,\ell) \in M} (\be_i - \be_{b_\ell(i)}).$$
Note that $\bx(t+\delta)$ is obtained by applying fractional local search to $\tilde{\bx}(t)$.
This cannot decrease the value of $F$, and hence
$$ F(\bx(t+\delta)) - F(\bx(t)) \geq F(\tilde{\bx}(t)) - F(\bx(t))
 = F\left(\bx(t) + \delta \sum_{(i,\ell) \in M} (\be_i - \be_{b_\ell(i)})\right) - F(\bx(t)).$$
Observe that up to first-order approximation, this increment is given by the
partial derivatives evaluated at $\bx(t)$. 
By Lemma~\ref{lem:Taylor}, the second-order term is proportional to $\delta^2$:
$$ F(\bx(t+\delta)) - F(\bx(t)) \geq \delta \sum_{(i,\ell) \in M} \left(\partdiff{F}{x_i}
 - \partdiff{F}{x_{b_\ell(i)}} \right) - n^2 \delta^2 \sup \Big|\mixdiff{F}{x_i}{x_j}\Big| $$
and from above,
$$ F(\bx(t+\delta)) - F(\bx(t)) \geq \frac{\delta}{1-t} \G(\bx(t))
%\left(\sum_{i \in C} (1-x_i) \partdiff{F}{x_i} - \sum_{j \notin C} x_j \partdiff{F}{x_j} \right)
  - n^2 \delta^2 \sup \Big|\mixdiff{F}{x_i}{x_j}\Big| $$
\end{proof}

%By taking the limit as $\delta \rightarrow 0$, the statement of Lemma~\ref{lem:temp-gain}
%becomes
%$$ (1-t) \frac{d}{dt} F(\bx(t)) \geq \sum_{i \in C} (1-x_i) \partdiff{F}{x_i}
 %- \sum_{j \notin C} x_j \partdiff{F}{x_j} $$
%and this differential equation is the basis of our analysis.
It remains to relate $\G(\bx(t))$ to the optimum (recall that $OPT = f(C)$),
using the complementary solutions found in Step 9.
In the next lemma, we show that $G(\bx)$ is lower bounded by the RHS of equation
\eqref{eq:localgains}.

\begin{lemma}
\label{lem:OPT-bound}
Assume $OPT=f(C)$, $\bx \in P_t(\cM)$,
$T_{\le\lambda}(\bx) = \{ i: x_i \leq \lambda \}$, and the value
of a local optimum on any of the subsets $T_{\le\lambda}(\bx)$ is at most $\beta$.
Then
$$ \G(\bx(t)) \geq OPT - 2 F(\bx) - 2 \beta t .$$
\end{lemma}

\begin{proof}
Submodularity means that partial derivatives can only decrease when coordinates
increase. Therefore by Lemma \ref{lem:submod-change},
$$ \T{1}{t}{1}{x} -\T{t}{t}{x}{x}  \leq \sum_{i \in C} (1-x_i) \partdiff{F}{x_i} \Big|_\bx $$
and similarly
$$ \T{t}{t}{x}{x} -\T{t}{0}{x}{0} \geq \sum_{j \notin C} x_j \partdiff{F}{x_j} \Big|_\bx.$$
Combining these inequalities, we obtain
\begin{equation}
\label{eq:vee-wedge}
2F(\bx(t)) + \G(\bx(t)) = 2~ \T{t}{t}{x}{x} + \sum_{i \in C} (1-x_i) \partdiff{F}{x_i}
 - \sum_{j \notin C} x_j \partdiff{F}{x_j} \geq \T{1}{t}{1}{x} + \T{t}{0}{x}{0}~.
\end{equation}
Let $A = \{i: x_i = t\}$ (and recall that $x_i \in [0,t]$ for all $i$). 
By applying the treshold lemma (see Lemma~\ref{lem:threshold} and the accompanying
example with equation \eqref{eq:threshold_hard}), we have:
%We apply the ``threshold lemma'' \ref{lem:threshold}:
% (see \cite{Vondrak09}, Lemma A.4): For a random threshold
%set $T_{> \lambda}(\bx) = \{i: x_i > \lambda \}$  where $\lambda \in [0,1]$ is uniformly random,
%$F(\bx) \geq \E{f(T_{>\lambda}(\bx))}$. Hence,
%$$ F(\bx \vee \b1_C) \geq \E{f(T_{> \lambda}(\bx \vee \b1_C))} = \E{f(T_{> \lambda}(\bx) \cup C)}.$$
%Note that since $x_i \in [0,t]$, we have $T_{>\lambda}(\bx) = \emptyset$ if $\lambda > t$
% (which happens with probability $1-t$). So we get:

\begin{equation}
\label{eq:vee-exp}
\T{1}{t}{1}{x}%\geq \E{f(T_{>\lambda}(\bx) \cup C)} 
\geq t~\E{ \, \TTR{1}{1}{1}{1}{0} \, \Big| \, \lambda < t} +  (1-t) \T{1}{0}{1}{0}
\end{equation}
By another application of Lemma~\ref{lem:threshold},
\begin{equation}
\label{eq:wedge-exp}
\T{t}{0}{x}{0} % \geq \E{f(T_{>\lambda}(\bx \wedge \b1_C))} = \E{f(T_{>\lambda}(\bx) \cap C)}
 \geq t~\E{ \, \TTL{1}{0}{1}{0}{0} \, \Big| \lambda < t}% \E{f(T_{>\lambda}(\bx) \cap C) \mid \lambda < t}.
\end{equation}
(We discarded the term conditioned on $\lambda \geq t$, where $T_{>\lambda}(\bx) = \emptyset$.)
It remains to combine this with a suitable set in the complement of $T_{>\lambda}(\bx)$.
Let $S_\kappa$ be a local optimum found inside $T_{\le\kappa}(\bx) = \overline{T_{>\lambda}(\bx)}$.
By Lemma 2.2 in \cite{LMNS09}, $f(S_\kappa)$ can be compared to any feasible subset
of $T_{\le\kappa}(\bx)$, e.g. $C_\kappa = C \cap T_{\le\kappa}(\bx)$, as follows:
$$ 2 f(S_\kappa) \geq f(S_\kappa \cup C_\kappa) + f(S_\kappa \cap C_\kappa)
 \geq f(S_\kappa \cup C_\kappa) = f(S_\kappa \cup (C \setminus T_{>\kappa}(\bx))).$$
We assume that $f(S_\kappa) \leq \beta$ for any $\kappa$.
Let us take expectation over $\lambda \in [0,1]$ uniformly random:
$$ 2 \beta \geq 2 \E{f(S_\lambda) \mid \lambda < t} \geq
%\TTR{
\E{f(S_\lambda \cup (C \setminus T_{>\lambda}(\bx))) \mid \lambda < t}.
$$
Now we can combine this with (\ref{eq:vee-exp}) and (\ref{eq:wedge-exp}):
\begin{eqnarray*}
 \T{1}{t}{1}{x} + \T{t}{0}{x}{0} + 2 \beta t & \geq &
 (1-t) \T{1}{0}{1}{0} + t~\E{ \, \TTL{1}{0}{1}{0}{0} + \TTR{1}{1}{1}{1}{0}
  + f(S_\lambda \cup (C \setminus T_{>\lambda}(\bx))) \, \Big| \, \lambda < t} \\
 %\E{f(T_{>\lambda}(\bx) \cup C)  + f(T_{>\lambda}(\bx) \cap C) + f(S_\lambda \cup (C \setminus T_{>\lambda}(\bx)))
%  \mid \lambda < t} \\
& \geq & (1-t) f(C) + t\left[~ \TTL{1}{0}{1}{0}{0} +  \TTL{0}{0}{0}{1}{0} ~\right]  \\
& \geq & (1-t) f(C) + t f(C) = f(C) = OPT.
\end{eqnarray*}
where the last two inequalities follow from submodularity. 
%since $(T_{>\lambda}(\bx) \cap C) \cup
 %(S_\lambda \cup (C \setminus T_{>\lambda}(\bx))) = S_\lambda \cup C$, and the third inequality
%follows from submodularity as $(T_{>\lambda}(\bx) \cup C) \cap (S_\lambda \cup C) = C$.
Together with (\ref{eq:vee-wedge}), this finishes the proof.
\end{proof}

%\begin{lemma}
%Let $\Phi(t) = F(\bx(t))$ and assume that the solution found in Step 4 of the algorithm
%is always at most $\beta$. For $\delta=o(1/n^2)$, the function $\Phi(t)$ satisfies
%$$ (1-t) \Phi'(t) + 2 \Phi(t) + 2 \beta t \geq (1-o(1)) OPT.$$
%\end{lemma}

\begin{proof}[Lemma \ref{lem:diffeq}]
By Lemma~\ref{lem:temp-gain} and \ref{lem:OPT-bound}, we get
$$
\frac{1-t}{\delta} (\Phi(t+\delta)-\Phi(t)) = \frac{1-t}{\delta} (F(\bx(t+\delta)) - F(\bx(t)))
 \geq OPT - 2 F(\bx) - 2 \beta t - n^2 \delta \, \sup \Big| \mixdiff{F}{x_i}{x_j} \Big|.$$
We have $|\mixdiff{F}{x_i}{x_j}| \leq 2 \max |f(S)| \leq 2n OPT$,
which implies the lemma.
\end{proof}

Now by taking $\delta \rightarrow 0$,
the statement of Lemma~\ref{lem:diffeq} leads naturally to the following differential equation:
$$ (1-t) \Phi'(t) \geq OPT - 2 \Phi(t) - 2t\beta.$$
This differential equation is very similar to the one we obtained in 
Section~\ref{sec:unconstrained}. Let us assume that $OPT=1$.
Starting from an initial point $F(t_0) = v_0$, the solution turns out to be
$$ \Phi(t) \geq \frac12 + \beta - 2 \beta t - \frac{(1-t)^2}{(1-t_0)^2}
 \left( \frac12 + \beta - 2 \beta t_0 - v_0 \right).$$
We start from initial conditions corresponding to the $0.309$-approximation
of \cite{Vondrak09}. It is proved in \cite{Vondrak09} that a fractional local optimum
at $t_0 = (1-t_0)^2 = \frac12 (3-\sqrt{5})$ has value $v_0 \geq \frac12 (1-t_0)
 \simeq 0.309$.
Therefore, we obtain the following solution for $t \geq \frac12 (3-\sqrt{5})$:
$$ \Phi(t) \geq \frac12 + \beta - 2 \beta t - (1-t)^2
 \left( \frac12 - 2 \beta + \frac{2 \beta}{3-\sqrt{5}} \right). $$
We solve for $\beta$ such that the maximum of the right-hand side equals $\beta$.
The solution is 
$$ \beta = \frac18 \left(2+\sqrt{5})(-5 + \sqrt{5} + \sqrt{-2 + 6 \sqrt{5}} \right).$$
Then, for some value of $t$ (which turns out to be roughly $0.53$),
we have $\Phi(t) \geq \beta$. It can be verified that $\beta > 0.325$;
this proves Theorem~\ref{thm:0.325-approx}.

%%%%%%%%%%%%%%%%%%%%%%%%%%%%%%%%
%%%%%%%%%%%%%%%%%%%%%%% APP: Hardness

\section{Additional Hardness Results}
\label{app:hardness}

\subsection{Matroid base constraints}

It is shown in \cite{Vondrak09} that it is hard to approximate submodular maximization
subject to a matroid base constraint with fractional base packing number $\nu = \ell/(\ell-1)$,
$\ell \in \ZZ$, better than $1/\ell$.
We showed in Theorem~\ref{thm:hard_matroidbase} that for $\ell=2$,
the threshold of $1/2$ can be improved to $1-e^{-1/2}$.
More generally, we show the following. 

\begin{theorem}
There exist instances of the problem $\max\{f(S):S\in \cB \}$,
such that a $(1-e^{-1/\ell}+\eps)$ approximation for any $\eps>0$ 
would require exponentially many value queries.
Here $f(S)$ is a nonnegative submodular function,
and $\cB$ is a collection of bases in a matroid with fractional base packing
number $\nu = \ell/(\ell-1), \ell \in \ZZ$. 
\end{theorem}

\begin{proof}
Let $\nu=\frac{\ell}{\ell-1}$.
Consider the hypergraph $H$ in Figure~\ref{fig:hardness_matroidbase},
with $\ell$ instead of $2$ hyperedges. Similarly let $A$ $(B)$ be the set of head (tail) vertices
respectively, and let the feasible sets be those that contain $\ell-1$ vertices of $A$ and
one vertex of $B$. (i.e. $\cB=\{S: |S\cap A|=\ell-1~\&~|S\cap B| = 1$).
The optimum can simply select the heads of the first $\ell-1$ hyperedges
and one of the tails of the last one, thus the value of $OPT=1$ remains unchanged.
On the other hand, $\overline{OPT}$ will decrease since the number of symmetric elements
has increased and there is a greater chance to miss a hyperedge.
Similar to the proof of Lemma~\ref{lem:symmetricgroups} and
Theorem~\ref{thm:hard_matroidbase} we obtain a unique symmetrized
vector $\bar{\bx}=(\frac1\ell,\frac1\ell,\ldots,\frac1\ell,
\frac{1}{k\ell},\frac{1}{k\ell},\ldots,\frac{1}{k\ell})$. Therefore,
$$
\gamma=\overline{OPT}=F(\bar{\bx})= \ell \left[ \frac{1}{\ell}
\left(1-\left(1-\frac{1}{k\ell}\right)^k\right)\right]
 \simeq 1-e^{-1/\ell},
$$
for sufficiently large $k$. Also it is easy to see that the feasible sets
of the refined instances, which are indeed the bases of a matroid,
are those that contain a $(\ell-1)/\ell$ fraction of the vertices in $A$
and $1/k\ell$ fraction of vertices in $B$. Therefore the fractional base packing number
of the refined instances is equal to $\frac{\ell}{\ell-1}$.
\end{proof}

\subsection{Matroid independence constraint}
\label{subsec:matroidindependence}

In this subsection we focus on the problem of maximizing a submodular
function subject to a matroid independence constraint.
Similar to Section~\ref{sec:hardness},
we construct our hard instances using directed hypergraphs.

\begin{theorem}
\label{thm:hard_matroidindependence}
There exist instances of the problem $\max\{f(S):S \in \mathcal{I}\}$
where $f$ is nonnegative submodular and $\cI$ are independent sets in a matroid
such that a $0.478$-approximation would require exponentially many value queries.
\end{theorem}

It is worth noting that the example we considered in Theorem \ref{thm:hard_matroidbase}
does not imply any hardness factor better than $1/2$ for the matroid independence problem.
The reason is that for the vector $\bar{\bx}=(0,0,\frac{1}{2k},\ldots,\frac{1}{2k})$, which is contained
in the matroid polytope $P(\cM)$, the value of the multilinear relaxation is $1/2$.
In other words, it is better for an algorithm not to select any vertex in the heads set $A$,
and try to select as much as possible from $B$.

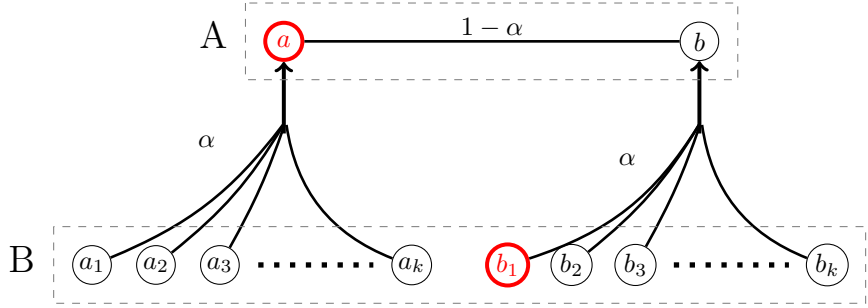
\begin{figure}
\centering
%\documentclass{article}
%\usepackage{fullpage,tikz}
%\usetikzlibrary{%
%decorations.fractals,%
%decorations.shapes,%
%decorations.text,%
%decorations.pathmorphing,%
%decorations.pathreplacing,%
%decorations.footprints,%
%decorations.markings}
%\pagestyle{empty}
%
%
%
%\oddsidemargin -.5in

\def\sl {3}
\def \ll {6.5}
\def\hi{3.5}
\def\rl{.6}
%\begin{document}

%\begin{figure}

%\begin{tabular}{cc}
\def \Pointsize {1pt}

\begin{tikzpicture}[inner sep=1.2pt,scale=.85,pre/.style={<-,shorten <=2pt,>=stealth,thick}, post/.style={->,shorten >=1pt,>=stealth,thick}]
\tikzstyle{every node}=[draw,circle];
%\tikzstyle{every node}=[draw,shape=circle,fill=black,inner sep=\Pointsize];
\path [line width=1.5pt,color=red,inner sep=2.6pt] (0,\hi) node (a) {$a$};
\path [inner sep=2.3pt] (\ll,\hi) node (b) {$b$};
\path  (a) +(-3,-\hi)  node (a1) {$a_1$};
\path (a) +(-2,-\hi)  node (a2) {$a_2$};
\path (a) +(-1,-\hi)  node (a3) {$a_3$};
\path (a) +(2,-\hi)  node (ak) {$a_k$};
\path [line width=1.5pt,color=red] (b) +(-3,-\hi)  node (b1) {$b_1$};
\path (b) +(-2,-\hi)  node (b2) {$b_2$};
\path (b) +(-1,-\hi)  node (b3) {$b_3$};
\path (b) +(2,-\hi)  node (bk) {$b_k$};

%\draw[->, >=latex, blue!20!white, line width=72pt] (4, 16) — node [black] {Optimization} +(-45:17cm);

\tikzstyle{every node}=[];
\path (a) +(0,-1.5) node (am) {} ;
\path (am) +(.03,.25) node (apm) {} ;
\path (b) +(0,-1.5 ) node (bm) {};
\path (bm) +(0.03,.25) node (bpm) {} ;

\draw [loosely dotted, line width=2pt] (a3) +(.6,0) -- +(2.4,0);
\draw [loosely dotted, line width=2pt] (b3) +(.6,0) -- +(2.4,0);

\path[->,line width=1.5pt] (am) edge (a)
	    (bm) edge  (b);
\path  [-,line width=1pt]    (a1) edge [bend right=15] node [above=20] {$\alpha$} (apm)
	     (a2) edge [bend right=10] (apm)
	     (a3) edge [bend right=5] (apm)
	    (ak) edge [bend left] (apm)
	   (b1) edge [bend right=20] node [above=15] {$\alpha$} (bpm)
	     (b2) edge [bend right=10] (bpm)
	     (b3) edge [bend right=5] (bpm)
	    (bk) edge [bend left]  (bpm);
\path [-, line width=1pt]   (a) edge [] node [above] {$1-\alpha$} (b);
\draw (a) +(-\rl,\rl) node (al){};
\draw (b) +(\rl,-\rl) node (br){};
\draw (bk) + (\rl,-\rl) node (bkr){};
\draw (a1) +(-\rl,\rl) node (a1l){};
\draw [color=gray,dashed] (al) rectangle (br)  (a1l)  rectangle (bkr);

\draw (al) +(-0.5,-.5) node {{\Large{A}}};
\draw (a1l) +(-0.5,-.5) node {{\Large{B}}};

\end{tikzpicture}
%\end{tabular}
%\end{figure} 

%\end{document}
%
\caption{Example for maximizing a submodular function subject to a matroid independence
constraint;  the hypergraph contains two directed hyperedges of weight $\alpha$
and the edge $(a,b)$ of weight $1-\alpha$; the constraint is that we pick
at most one vertex from each of $A$ and $B$.}
\label{fig:hardness_matroidindependence}
\end{figure}

\medskip
\noindent{\bf Instance 2.} To resolve this issue,
we perturb the instance by adding an undirected edge $(a,b)$ of weight $1-\alpha$
and we decrease the weight of the hyperedges to $\alpha$, where the value of $\alpha$
will be optimized later (see Figure~\ref{fig:hardness_matroidindependence}).
The objective function is again the (directed) cut function, where the edge $(a,b)$
contributes $1-\alpha$ if we pick exactly one of vertices $a,b$.
Therefore the value of the optimum remains unchanged, $OPT=\alpha + (1-\alpha)=1$.
On the other hand the optimal symmetrized vector $\bar{\bx}$ should have a non-zero
value for the head vertices, otherwise the edge $(a,b)$ would not have any
contribution to $F(\bar{\bx})$.\\

\begin{proof}[Theorem \ref{thm:hard_matroidindependence}]
Let $H$ be the hypergraph of Figure~\ref{fig:hardness_matroidindependence},
and consider the problem $\max\{f(S): S\in \cI\}$, where $f$ is the cut function of $H$
and $\cI$ is the set of independent sets of the matroid $\cM_{A,B}$ defined in subsection
\ref{subsec:matroidindependence}. 
Observe that Lemma \ref{lem:symmetricgroups} can be applied to our instance  as well,
thus we may use equation \eqref{eq:symmetrization} to obtain the symmetrized
vectors $\bar{\bx}$.
Moreover, the matroid polytope can be described by the following equations:
\begin{equation}
\label{eq:indproperty}
x\in P(\cM_{A,B}) \Leftrightarrow 
\begin{cases} x_a+x_b\leq 1\\
\sum_{i=1}^k (x_{a_i}+x_{b_i})\leq1.
\end{cases}
\end{equation}
Since the vertices of the set $B$ only contribute as tails of hyperedges,
the value of $F(\bar{\bx})$ can only increase if we increase the value of $\bar{\bx}$ on
the vertices in $B$. Therefore, we can assume (using equations \eqref{eq:symmetrization} and
\eqref{eq:indproperty}) that
\begin{eqnarray*}
& \bar{x}_a=\bar{x}_b \leq \frac{1}{2} & \\
& \bar{x}_{a_1}=\bar{x}_{b_1}=\ldots=\bar{x}_{a_k}=\bar{x}_{b_k}=\frac{1}{2k}. &
\end{eqnarray*}
Let $\bar{x}_a = q$; we may compute the value of $\overline{OPT}$ as follows:
$$
\overline{OPT}=F(\bar{\bx})=2\alpha\left[(1-q) \left(1-(1-\frac{1}{k})^k\right)\right]
 + (1-\alpha)\left[2q(1-q)\right],
$$
where $q\leq 1/2$.
By optimizing numerically over $\alpha$, we find that the smallest value of $\overline{OPT}$
is obtained when $\alpha \simeq 0.3513$. In this case we have $\gamma=\overline{OPT}
 \simeq 0.4773$. Also, similarly to Theorem \ref{thm:hard_matroidbase},
the refined instances are in fact instances of a submodular maximization 
problem over independent sets of a matroid (a partition matroid whose ground set
is partitioned into $A\cup B$ and we have to take at most half of the elements of $A$
and $1/2k$ fraction of elements in $B$).
\end{proof}

\subsection{Cardinality constraint}

Although we do not know how to improve the hardness of approximating general submodular
functions without any additional constraint to a value smaller than $1/2$, we can show
that adding a simple cardinality constraint makes the problem harder. In particular,
we show that it is hard to approximate a submodular function subject to a cardinality
constraint within a factor of $0.491$. 

\begin{corollary}
There exist instances of the problem $\max\{f(S):|S| \leq \ell\}$ with $f$ nonnegative submodular
such that a $0.491$-approximation would require exponentially many value queries.
\end{corollary} 

We remark that a related problem, $\max \{f(S): |S|=k\}$, is at least as difficult
to approximate: we can reduce $\max \{f(S): |S| \leq \ell\}$ to it by trying
all possible values $k=0,1,2,\ldots,\ell$.

\medskip

\begin{proof}
Let $\ell=2$, and let $H$ be the hypergraph we considered in previous theorem
and $f$ be the cut function of $H$. Similar to the proof of 
Theorem~\ref{thm:hard_matroidindependence}, we have $OPT=1$ and we may use
equation \eqref{eq:symmetrization} to obtain the value of $\bar{\bx}$.
In this case the feasibility polytope will be
\begin{equation}
\label{eq:cardinalproperty}
x\in P(|S|\leq 2) \Leftrightarrow 
 x_a+x_b+\sum_{i=1}^k (x_{a_i}+x_{b_i})\leq 2,
\end{equation}
however, we may assume that we have equality for the maximum value of $F(\bar{\bx})$,
otherwise we can simply increase the $\bar{\bx}$ value of the tail vertices in $B$
and this can only increase $F(\bar{\bx})$. Let $\bar{x}_a=q$ and $x_{a_1}=p$ and $z=kp$.
Using equations \eqref{eq:symmetrization} and \eqref{eq:cardinalproperty} we have
$$
2q + 2kp=2 \Rightarrow kp=z=1-q.
$$
Finally, we can compute the value of $\overline{OPT}$:
\begin{eqnarray*}
\overline{OPT}=F(\bar{\bx}) & = & 2\alpha\left[(1-q) \left(1-(1-p)^k\right)\right]
 + (1-\alpha)\left[2q(1-q)\right]\\
 & = & 2\alpha z(1-e^{-z}) + 2(1-\alpha)z(1-z).
\end{eqnarray*}
Again by optimizing over $\alpha$, the smallest value of $\overline{OPT}$
is obtained when $\alpha \simeq 0.15$. In this case we have $\gamma \simeq 0.49098$.
The refined instances are instances of submodular maximization
subject to a cardinality constraint, where the constraint is to choose at most
$\frac{1}{k+1}$ fraction of the all the elements in the ground set.
\end{proof}

\end{document}